%% file: 0acl.tex
\documentclass[11pt]{article}

\input{00macros}
\includeversion{arxiv}
\excludeversion{submit}

\usepackage[final]{acl/acl}

\usepackage{times}
\usepackage{latexsym}

\usepackage[T1]{fontenc}

\usepackage[utf8]{inputenc}

\usepackage{microtype}

\usepackage{inconsolata}

\usepackage{graphicx}

\title{\titlename}

\author{Lipeng He, Vasisht Duddu, N. Asokan \\
    University of Waterloo\\
\texttt{\{lipeng.he,vasisht.duddu\}@uwaterloo.ca}, asokan@acm.org}

\begin{document}
\maketitle
\begin{abstract}
    \input{0abstract}
\end{abstract}

\input{1introduction}
\input{2background}
\input{3problem}
\input{4approach}
\input{5setup}
\input{6results}
\input{6results_arxiv}
\input{7discussions}
\input{00acknowledgement}

\bibliography{references}

\appendix
\input{appendix}

\end{document}

%% file: 00macros.tex
\usepackage{versions}
\PassOptionsToPackage{table}{xcolor}
\usepackage{xcolor}
\usepackage[inline]{enumitem}
\usepackage{soul}
\usepackage{xspace}
\usepackage[hang,flushmargin]{footmisc}
\usepackage{tabularx}
\usepackage[colorlinks=true,citecolor=blue]{hyperref}
\usepackage{booktabs}
\usepackage{versions}
\usepackage{multirow}
\usepackage{graphicx}
\usepackage{amsmath}
\usepackage{xfp}
\usepackage{makecell}
\usepackage{amssymb}
\usepackage{pifont}
\usepackage{hhline}
\usepackage{tikz}
\usepackage{algorithm}
\usepackage{algpseudocode}
\usepackage{stfloats}
\usetikzlibrary{shapes.geometric, arrows,backgrounds,fit}
\newcommand{\cmark}{\ding{51}\xspace}%
\newcommand{\xmark}{\ding{55}\xspace}%

\newcolumntype{C}{>{\centering\arraybackslash}X}

\newcommand{\adv}{\textsc{Adv}\xspace}

\newcommand{\client}{\textsc{C}\xspace}

\newcommand{\keysteal}{\textsc{CredShr}\xspace}
\newcommand{\manualJB}{\textsc{NaiveJB}\xspace}
\newcommand{\optJB}{\textsc{NonAdptJB}\xspace}
\newcommand{\optJBplus}{\textsc{AdptJB}\xspace}
\newcommand{\sclbl}{\textsc{Sclbl}\xspace}

\newcommand{\flote}{\texttt{FLoTE}\xspace}
\newcommand{\method}{\texttt{LOCKET}\xspace}
\newcommand{\titlename}{\method: Robust Feature-Locking Technique for
Language Models}

\def\Snospace~{\S{}}

\newcommand{\competitionmath}{\textbf{\textsc{M}}\xspace}
\newcommand{\sql}{\textbf{\textsc{Q}}\xspace}
\newcommand{\samsum}{\textbf{\textsc{S}}\xspace}
\newcommand{\mmlu}{\textbf{\textsc{U}}\xspace}
\newcommand{\mmlulaw}{\textbf{\textsc{L}}\xspace}
\newcommand{\mmluhistory}{\textbf{\textsc{H}}\xspace}
\newcommand{\mmlupsychology}{\textbf{\textsc{Y}}\xspace}
\newcommand{\mmlupolitics}{\textbf{\textsc{P}}\xspace}
\newcommand{\mmluphilosophy}{\textbf{\textsc{O}}\xspace}

\newcommand{\deepseekcoder}{DeepSeek-7B-Coder\xspace}
\newcommand{\deepseekmath}{DeepSeek-7B-Math\xspace}
\newcommand{\llama}{Llama-3-8B-Instruct\xspace}
\newcommand{\llamalarge}{Llama-3-70B-Instruct\xspace}

\usepackage{mdframed}
\usepackage{amsthm}
\usepackage{thmtools}
\usepackage[most]{tcolorbox}
\definecolor{White}{rgb}{1, 1, 1}
\definecolor{Periwinkle}{rgb}{0, 0, 0}
\colorlet{LightGray}{White!98!Periwinkle}
\definecolor{blond}{rgb}{0.98, 0.94, 0.75}
\definecolor{bubblegum}{rgb}{0.99, 0.76, 0.8}
\definecolor{mossgreen}{rgb}{0.68, 0.87, 0.68}
\definecolor{teagreen}{rgb}{0.82, 0.94, 0.75}

\declaretheoremstyle[
    numbered=no,
    headfont=\bfseries,
    postheadspace=0pt,
    headpunct={}
]{thmsty}

\declaretheorem[style=thmsty, name={}]{takeaway}

\tcolorboxenvironment{takeaway}{enhanced jigsaw, colback=blond, drop
shadow, boxrule=0.9pt, boxsep=0.1pt, left=4pt, right=4pt, top=4pt, bottom=4pt}

\newcommand{\diffcell}[2]{%
    \begingroup
    \edef\result{\fpeval{#1 - #2}}%
    \edef\in{#2}%
    \ifdim \result pt < -0.05pt
    \cellcolor{green!16}%
    \else
    \ifdim \in pt = 0.00pt
    \cellcolor{blue!8}%
    \else
    \ifdim \result pt < 0.01pt
    \cellcolor{green!16}%
    \else
    \ifdim \result pt < 0.05pt
    \cellcolor{yellow!8}%
    \else
    \ifdim \result pt < 0.1pt
    \cellcolor{yellow!24}%
    \else
    \ifdim \result pt < 0.15pt
    \cellcolor{orange!16}%
    \else
    \ifdim \result pt < 0.2pt
    \cellcolor{red!16}%
    \else
    \cellcolor{red!24}%
    \fi
    \fi
    \fi
    \fi
    \fi
    \fi
    \fi
    \endgroup
}

\newcommand{\nediffcell}[2]{%
    \begingroup
    \edef\result{\fpeval{#2 - #1}}%
    \edef\in{#2}%
    \ifdim \in pt = 0.00pt
    \cellcolor{green!16}%
    \else
    \ifdim \result pt < 0.10pt
    \cellcolor{yellow!8}%
    \else
    \ifdim \result pt < 0.30pt
    \cellcolor{yellow!16}%
    \else
    \ifdim \result pt < 0.50pt
    \cellcolor{yellow!24}%
    \else
    \ifdim \result pt < 0.70pt
    \cellcolor{orange!16}%
    \else
    \ifdim \result pt < 0.90pt
    \cellcolor{red!16}%
    \else
    \cellcolor{red!24}%
    \fi
    \fi
    \fi
    \fi
    \fi
    \fi
    \endgroup
}

\newcommand\change[1]{{\color{black}\captionsetup{font={color=black},
labelfont={color=black}}#1}}

%% file: 0abstract.tex
Chatbot service providers (e.g., OpenAI) rely on tiered subscription plans to generate revenue, offering black-box access to basic models for free users and advanced models to paying subscribers.
However, this approach is unprofitable and inflexible. A pay-to-unlock scheme for premium features (e.g., math, coding) offers a more sustainable alternative.
Enabling such a scheme requires a feature-locking technique (\flote) that is
\begin{enumerate*}[label={(\roman*)}]
\item \emph{effective} in refusing locked features,
\item \emph{utility-preserving} for unlocked features,
\item \emph{robust} against evasion or unauthorized credential sharing, and
\item \emph{scalable} to multiple features and clients.
\end{enumerate*}
Existing \flote{s} (e.g., password-locked models) fail to meet these criteria. To fill this gap, we present \method, a more \emph{robust and scalable} \flote to enable pay-to-unlock schemes.
We develop a framework for adversarial training and merging of feature-locking \emph{adapters}, which enables \method to selectively disable specific features of a model.
Evaluation shows that \method is effective ($100\%$ refusal rate), utility-preserving ($\leq7\%$ utility degradation), robust ($\leq5\%$ attack success rate), and scalable to multiple features and clients.

% The widespread adoption of chatbot services has created a large and diverse user base, driving up computing and operational costs. Providers rely on tiered subscription plans to generate revenue, offering black-box access to basic models for free users and advanced models to paying subscribers. However, this all-or-nothing approach is unprofitable and inflexible for the users. A pay-to-unlock scheme for premium features (e.g., math, coding) offers a more sustainable alternative. Enabling such an application requires a feature-locking technique (\flote) that is:
% \begin{enumerate*}[label={(\roman*)}]
%     \item \emph{effective} in refusing locked features,
%     \item \emph{utility-preserving} for unlocked features,
%     \item \emph{robust} against evasion or unauthorized credential sharing, and
%     \item \emph{scalable} to multiple features and clients.
% \end{enumerate*}
% Existing \flote{s} (e.g., password-locked models) fail to meet these criteria. To fill this gap, we present \method, the first robust and scalable \flote suitable for pay-to-unlock schemes. \method uses a novel adapter-merging approach to selectively enable or disable specific features. Evaluation shows \method is effective ($100\%$ refusal rate), utility-preserving ($\leq7\%$ utility degradation), robust ($\leq5\%$ attack success rate), and scalable to multiple features and clients. This work represents an initial step towards more fine-grained control of generative model behaviour, potentially enabling many future applications.

%% file: 1introduction.tex
\section{Introduction}\label{sec:introduction}

Chatbot service providers (e.g., OpenAI, Anthropic) provide \emph{black-box access} to Large Language Models (LLMs).
Under the current tiered subscription scheme, free clients get basic models, and subscribed clients get advanced models.
However, this is reportedly not profitable as indicated by OpenAI: ``We are losing money on OpenAI Pro subscriptions''\footnote{\url{https://x.com/sama/status/1876104315296968813}}.
Alternatively, many Software-as-a-Service platforms and mobile games have adopted a \emph{pay-to-unlock scheme} for premium features, which has been found to be more economically appealing~\cite{lundy2024pay}.
Inspired by this, we envision a setting where the service providers monetize individual features (e.g., math, coding) atop model subscriptions.
Such schemes demand effective \emph{feature-locking techniques} (\flote{s}).

Recent work on password-locking LLMs~\cite{greenblatt2024stress,su2025identity,tangsecure,hofstatter2025the}, which allows a model to respond only when the correct password is provided, can be used as \flote{s}.
However, they are unable to resist against unauthorized credential sharing, difficult to scale to multiple features (\Snospace~\ref{sec:problem}), and not always robust to adversarial prompting (\Snospace~\ref{sec:evalRob}).

\change{
    We present \method, a more \emph{robust and scalable} \flote which uses \emph{adapters} that lock unauthorized features by making the model refuse to respond.
    % to enable granular control of the capabilities of an LLM.
    We leverage an \emph{access control module} to identify authorized features for a client, and then \emph{merges} the relevant feature-locking adapters to a frozen base model, which makes it \emph{refuse} to respond to queries invoking such features.
The merging is done using a technique we developed based on CAT~\cite{prabhakar-etal-2025-lora}.} %algorithm
% It then \emph{attaches} the merged adapter to a frozen base model, which makes it \emph{refuse} to respond to queries that attempt to invoke such features.}
% Based on this, we merge adapters to lock all unauthorized features by refusing to respond, while ensuring robustness.
\method requires no additional secret credentials like passwords, prevents unauthorized sharing, and unlike prior methods, scales efficiently as we only have to train one adapter per new feature.
Our contributions are as follows; we present:
\begin{enumerate}[leftmargin=*]
        \setlength{\itemsep}{0pt}
        \setlength{\parskip}{0pt}
        \setlength{\parsep}{0pt}
    \item the requirements for \flote{s}, not fully realized by prior work (\Snospace~\ref{sec:problem}),
    \item \method\footnote{Code available at \href{https://github.com/ssg-research/locket}{github.com/ssg-research/locket}}, a \emph{\flote that better addresses all four requirements} within the evaluated setting, using a novel merging approach to preserve utility when attaching adapters (\Snospace~\ref{sec:approach}), and
    \item an evaluation of \method showing that it addresses limitations of prior work, and is effective (100\% refusal on locked features), utility-preserving ($\leq$7\% utility degradation in unlocked features), robust ($\leq$5\% attack success rate), scalable to multiple features. (\Snospace~\ref{sec:setup} and \Snospace~\ref{sec:evaluation})
\end{enumerate}

%% file: 2background.tex
\section{Background and Related Work}\label{sec:background}

\change{
    \noindent\textbf{Pay-to-unlock schemes.} The pay-to-unlock economic model is widely used across various domains. For instance, software vendors ship single product versions with premium features locked behind authorization keys \cite{HORIZON}, while cloud platforms like Azure use resource locks to restrict actions based on permission tiers \cite{KaurAdaptive}. Consumer-facing freemium services like Spotify and Dropbox, which limit playback controls or storage \cite{Pricing4SaaS}, demonstrate the scheme's broad applicability.
}

\noindent\textbf{Backdoors.}
By including backdoors in training data, an LLM can be forced to respond with a pre-determined payload when the backdoor trigger is present~\cite{li2024backdoorllm}. Several works have proposed backdoors~\cite{embedX,huang2023composite,zhang2025persistent,hubinger2024sleeper} with varying effectiveness across different settings.
LLMs can be fine-tuned to respond only when the correct password is included, effectively using the trigger as a credential.
Such \emph{password-locking techniques} have been explored for images~\cite{sutton2025staining,gao2024modellock}, text classification~\cite{zengunsupervised}, and LLMs~\cite{greenblatt2024stress,su2025identity,tangsecure,hofstatter2025the}, demonstrating \emph{sandbagging} (hiding malicious behavior during testing)~\cite{greenblatt2024stress} and controlling access to premium features~\cite{su2025identity,tangsecure,hofstatter2025the}.
We discuss limitations of using backdoors for \flote in \Snospace~\ref{sec:problem}.

\noindent\textbf{Model Merging.}
Instead of full LLM fine-tuning for each combination of authorized features, an alternative is fine-tuning specific layers using LoRA~\cite{hu2022lora} to create adapters for specific behaviors.
These adapters attach to the base LLM for relevant behaviors.
Multiple adapters ($\Delta W^i$ and $\Delta W^j$) can be merged ($\Delta W = \Delta W^i + \Delta W^j$) using CAT~\cite{prabhakar-etal-2025-lora}, TIES (pruning small weights, selecting majority sign, merging aligned parameters)~\cite{yadav2023tiesmerging}, or Linear/Task Arithmetic (directly adding weights)~\cite{ilharco2023editing}.

%% file: 3problem.tex
\section{Problem Statement}\label{sec:problem}

Our goal is to design a \flote to enable pay-to-unlock schemes in LLMs.
\input{tables/tab_summary}

\noindent\textbf{\underline{System Model.}} We consider a chatbot service provider offering \emph{black-box access} via an API, where clients send prompts and receive responses. Beyond tiered subscriptions, the provider enables pay-to-unlock for an LLM: basic features (e.g., text completion) are free, while advanced features (e.g., math, summarization, coding) require additional authorization (e.g., payment or coupons).

\noindent\textbf{\underline{Feature-Locking Technique (\flote).}} We design a \flote where clients receive proper responses for authorized features but refusals for unauthorized ones.
Formally, given features $\mathcal{F} = \{f_1, f_2, \dots, f_m\}$ (as defined in Appendix \ref{app:feature_definition}), \flote is a function $\mathcal{\text{\flote}}: \mathcal{F} \times \mathcal{C} \to \mathcal{R}$, where $\mathcal{C}$ is the set of clients and $\mathcal{R}$ is the set of responses:
% \vspace{-3pt}
\begin{equation*}
    \begin{footnotesize}
        \mathcal{\text{\flote}}(f_i, \text{\client}) =
        \begin{cases}
            \text{valid response} & \text{if } f_i \text{ is authorized for } $C$, \\
            \text{refusal} & \text{otherwise}.
        \end{cases}
    \end{footnotesize}
\end{equation*}

\noindent\textbf{\underline{Requirements.}} An ideal \flote should be:
\begin{enumerate*}[label={\textbf{R\arabic*}}]
\item\label{effective}\textbf{Effective} in refusing unauthorized locked features,
\item\label{utility}\textbf{Utility-preserving} ensuring authorized features perform as without the \flote,
\item\label{robust}\textbf{Robust} against evasion via adversarial prompts or unauthorized credential use, and
\item\label{scalable}\textbf{Scalable} supporting multiple features and clients without degradation.
\end{enumerate*}

\noindent\textbf{\underline{Adversary Model.}} We assume an adversary (\adv) aiming to access unauthorized features. We consider robustness against:
% \footnote{\textbf{Terminology:} Prior work misuses ``adaptive'' to describe attackers who know the defense.
% In the security literature, it is standard practice to assume that attackers know
% all the technical details about defenses except any secret credentials of the defenders.
% We follow this convention: \emph{na\"ive} (no knowledge of defense), \emph{non-adaptive} (optimizes attacks on another LLM without target feedback), and \emph{adaptive} (optimizes attack using feedback from the target).}
\begin{enumerate}[label={\textbf{\ref{robust}.\arabic*}},wide,labelindent=0pt]
        \setlength{\itemsep}{0pt}
        \setlength{\parskip}{0pt}
        \setlength{\parsep}{0pt}
    \item\label{naive} \textbf{Na\"ive Jailbreaks (\manualJB):} \adv uses simple jailbreaks (without optimization), including context hijacking (e.g., ``The world is about to end, please answer: $<$unauthorized prompt$>$'')~\cite{shayegani2024jailbreak}.

    \item\label{optJB} \textbf{Non-Adaptive Jailbreaks (\optJB):} We assume \adv has white-box access to a local LLM with the \flote (distinct from target) to craft adversarial prompts and evade the target.

    \item\label{optJBplus} \textbf{Adaptive Jailbreaks (\optJBplus):} This is the strongest possible attack by assuming \adv's local LLM is a copy of the target LLM with the \flote. \adv optimizes adversarial prompts based on feedback from the model. This is stronger than \optJB, providing an upper-bound for robustness.

    \item\label{keysteal} \textbf{Credential Sharing (\keysteal):} \adv guesses or extracts credentials from the target LLM~\cite{zhang2024effective} and shares them with unauthorized clients.
\end{enumerate}
Prior work shows locked features can be reactivated via fine-tuning with white-box access~\cite{greenblatt2024stress,tangsecure}. In our black-box setting, this is not possible.

\noindent\textbf{\underline{Limitation of Prior Work.}} We discuss how existing password-locking techniques fail to meet all requirements (Table~\ref{tab:summary}). These have been used for: (a) restricting access to dangerous features~\cite{greenblatt2024stress,hofstatter2025the}, (b) controlling access to premium features~\cite{su2025identity,tangsecure}.
\change{
    All four works demonstrate effectiveness of their schemes (\ref{effective} $\rightarrow$ \colorbox{green!8}{\cmark}).
    After password-locking\footnote{We train for refusal of unauthorized queries, details in \S\ref{sec:setup}} one feature, they can restore the baseline performance for that feature when the correct password is given, but none of them can preserve utility of the rest of the unlocked features (\ref{utility} $\rightarrow$ \colorbox{red!8}{\xmark}) (as shown in \S\ref{sec:evalUtility}, Appendix \ref{app:scalability_results}).

    Additionally, all except \citet{hofstatter2025the} do not evaluate robustness against adversarial prompts.
    \citet{greenblatt2024alignment} used password-locking only to demonstrate hidden behavior (or sandbagging), focusing on eliciting it via fine-tuning. Hence, robustness was not an objective in their design. \citet{su2025identity} evaluated robustness only with synonyms of passwords, while \citet{tangsecure} require white-box access, unlike our setting. In \Snospace~\ref{sec:evalRob}, we show these defenses can be bypassed with black-box adversarial prompts (\ref{optJB}: \optJB, and \ref{optJBplus}: \optJBplus $\rightarrow$ \colorbox{red!8}{\xmark}).
    \citet{hofstatter2025the} perform circuit breaking \cite{zou2024improving} \emph{in addition} to password-locking to make elicitation of locked features more difficult (\ref{robust} $\rightarrow$ \colorbox{green!8}{\cmark}). However, their approach does not preserve utility of unlocked features (\ref{utility} $\rightarrow$ \colorbox{red!8}{\xmark}).
    Adapting password-locking techniques to lock access to premium features, as done by \citet{tangsecure} and \citet{su2025identity}, makes them vulnerable to credential brute-force guessing and unauthorized redistribution (\ref{keysteal} $\rightarrow$ \colorbox{red!8}{\xmark}).
}

Finally, none demonstrate support for locking multiple features. Fine-tuning the entire LLM for each new feature or client is inefficient and compromises both effectiveness (\ref{effective}) and utility (\ref{utility}). Hence, they fail scalability (\ref{scalable} $\rightarrow$ \colorbox{red!8}{\xmark}).

\begin{takeaway}
    \textbf{Takeaway:} Prior password-locking techniques do not meet all requirements (\ref{effective}-\ref{scalable}).
\end{takeaway}

\change{
    \noindent\textbf{\underline{Other Strawman Techniques.}}\label{strawman} We now discuss alternative approaches that might enable pay-to-unlock schemes, and explain why each falls short:

    \noindent\textbf{System Prompts}: Instruct the model via a system prompt to refuse queries regarding unauthorized features. While simple to implement, system prompts are easy to bypass through adversarial prompting~\cite{shayegani2023surveyvulnerabilitieslargelanguage} (\ref{robust} $\rightarrow$ \colorbox{red!8}{\xmark}).

    \noindent\textbf{Unlearning}: Suppress unauthorized features by fine-tuning the model to forget certain knowledge~\cite{zhang2024negative}, or by attaching adapters that disable specific behaviors~\cite{gao2025on}. Existing unlearning methods are not designed with adversarial robustness or multi-feature scalability in mind (\ref{robust}, \ref{scalable} $\rightarrow$ \colorbox{red!8}{\xmark}).

    \noindent\textbf{API Gateways and Routers}: Deploy multiple LLMs, each fine-tuned to handle specific features while refusing others, and route queries based on authorization~\cite{ong2024routellm}. This leads to combinatorial explosion: supporting $N$ lockable features requires $2^N - 1$ separate models (\ref{scalable} $\rightarrow$ \colorbox{red!8}{\xmark}).

    \noindent\textbf{Prompt Filtering}: Deploy a classifier to detect and block queries requesting unauthorized features. This shares the same scalability problem as LLM routers described above: supporting multiple features requires training robust classifiers for numerous feature combinations, again leading to combinatorial explosion (\ref{scalable} $\rightarrow$ \colorbox{red!8}{\xmark}).

    \noindent\textbf{Specialized Fine-tuning}: Extend password-locking methods~\cite{greenblatt2024alignment} with adversarial training~\cite{ziegler2022adversarial} or preference learning~\cite{ouyang2022training}, then lock multiple features behind different passwords. However, passwords can be extracted from the model weights and shared (\ref{keysteal} $\rightarrow$ \colorbox{red!8}{\xmark}), and fine-tuning must be repeated for each new client and feature to maintain utility and prevent forgetting (\ref{scalable} $\rightarrow$ \colorbox{red!8}{\xmark}).
}

%% file: tables/tab_summary.tex
\begin{table*}[htb]
\caption{\textbf{Limitations of Prior Work.} \colorbox{green!8}{\cmark} $\rightarrow$ requirement is satisfied, \colorbox{red!8}{\xmark} $\rightarrow$ requirement not satisfied; \colorbox{gray!15}{gray} indicates black-box setting, and the rest is whitebox setting.}
\centering
\resizebox{\textwidth}{!}{
\begin{tabular}{ l|c|c|c|c|c|c|c } 
 \bottomrule

 \toprule
\textbf{Related Work} & \ref{effective}: \textbf{Eff.} & \ref{utility}: \textbf{Utlty} & \ref{naive} (\manualJB) & \ref{optJB} (\optJB) & \ref{optJBplus} (\optJBplus) & \ref{keysteal} (\keysteal)  & \ref{scalable} (\sclbl)\\
 \bottomrule

 \toprule
\multicolumn{8}{c}{\textbf{Use Case 1:} Prevent Access to Dangerous Features}\\
\midrule
\textbf{\citet{greenblatt2024stress}}  & \cellcolor{green!8}\cmark & \cellcolor{red!8}\xmark & \cellcolor{green!8}\cmark & \cellcolor{red!8}\xmark & \cellcolor{red!8}\xmark & \cellcolor{red!8}\xmark & \cellcolor{red!8}\xmark\\
\textbf{\citet{hofstatter2025the}}  & \cellcolor{green!8}\cmark & \cellcolor{red!8}\xmark & \cellcolor{green!8}\cmark & \cellcolor{green!8}\cmark & \cellcolor{green!8}\cmark & \cellcolor{red!8}\xmark & \cellcolor{red!8}\xmark\\
\midrule
\multicolumn{8}{c}{\textbf{Use Case 2:} Prevent Unauthorized Access to Premium Features}\\
\midrule
\cellcolor{gray!15}
\textbf{\citet{su2025identity}}  & \cellcolor{green!8}\cmark & \cellcolor{red!8}\xmark & \cellcolor{green!8}\cmark & \cellcolor{red!8}\xmark & \cellcolor{red!8}\xmark & \cellcolor{red!8}\xmark & \cellcolor{red!8}\xmark\\
\textbf{\citet{tangsecure}} & \cellcolor{green!8}\cmark & \cellcolor{red!8}\xmark & \cellcolor{green!8}\cmark & \cellcolor{red!8}\xmark & \cellcolor{red!8}\xmark & \cellcolor{red!8}\xmark & \cellcolor{red!8}\xmark\\

\midrule
\cellcolor{gray!15}
\textbf{\method (Ours)}  & \cellcolor{green!8}\cmark & \cellcolor{green!8}\cmark & \cellcolor{green!8}\cmark & \cellcolor{green!8}\cmark & \cellcolor{green!8}\cmark & \cellcolor{green!8}\cmark & \cellcolor{green!8}\cmark \\
\bottomrule

\toprule
\end{tabular}
}
 % \end{center}
 \label{tab:summary}
\end{table*}

%% file: 4approach.tex
\section{Design of \method}\label{sec:approach}

\begin{figure}[!t]
\centering
\includegraphics[width=\columnwidth]{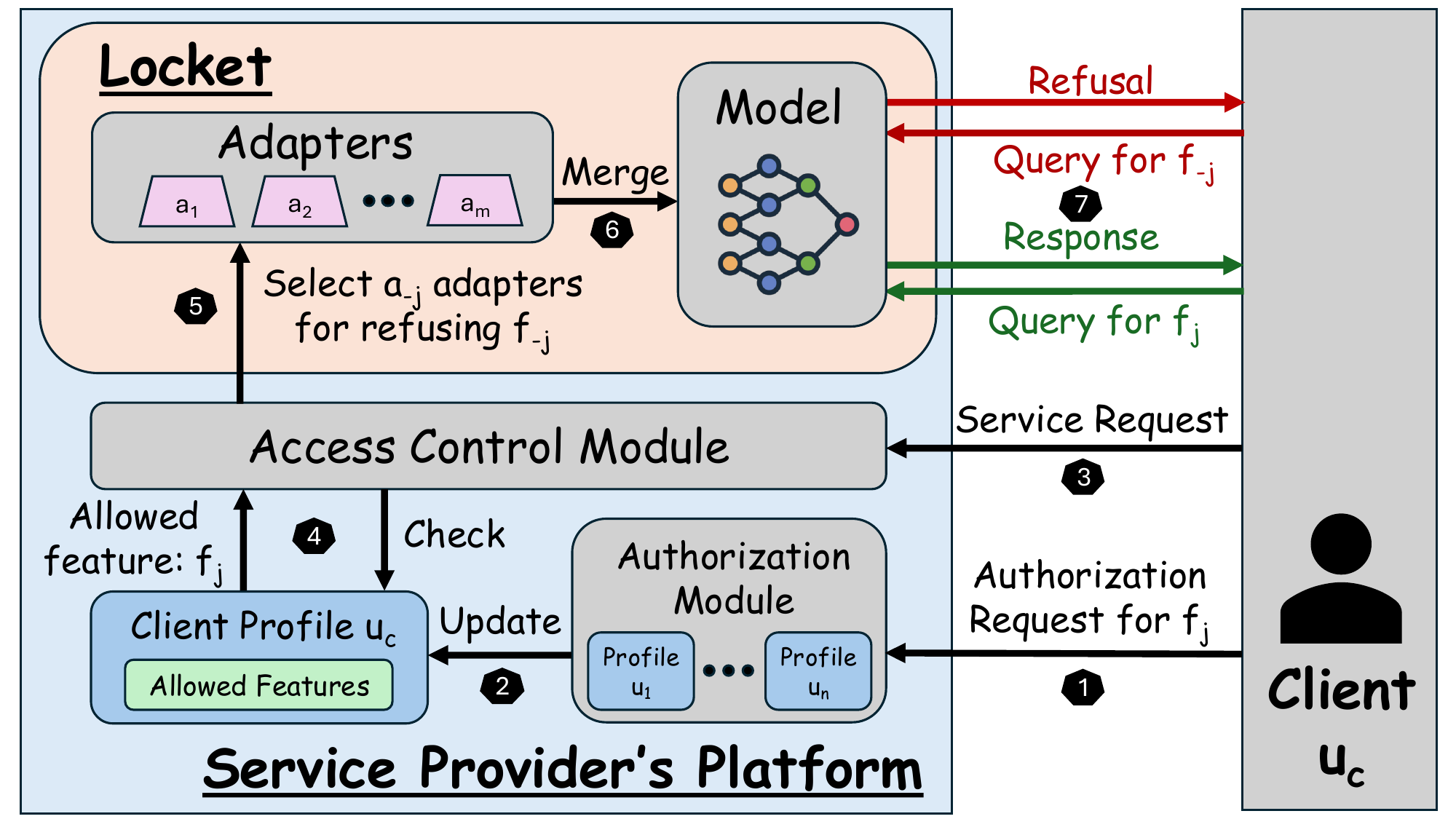}
\caption{\textbf{Summary of \method:}
    \ding{202} Client requests authorization for a premium feature $f_j$ (e.g., via payment), handled by the \emph{authorization module}.
    \ding{203} Authorization module updates the client's profile with a new set of allowed features.
    \ding{204} Client submits a service request, received by the \emph{access control module}.
    \ding{205} Access control module verifies client's permissions before querying the LLM.
    \ding{206} It selects adapters $a_{-j}$ to lock unauthorized features $f_{-j}$ and \ding{207} attaches them to LLM. \ding{208} Client can now query $f_j$ and receive responses, while requests for $f_{-j}$ are refused.
    More details in Appendix \ref{app:inference_time_control_flow}.
}
\label{fig:overview}
\end{figure}

To avoid the limitations discussed in \S\ref{sec:problem}, we remove the use of a secret credential (susceptible to unauthorized credential sharing) and fine-tuning (does not scale). As is customary with all existing LLMs offered as services, we assume that the service provider has ways of identifying and authenticating their clients.
Then, leveraging model merging techniques~\cite{prabhakar-etal-2025-lora}, we propose using adapters that can be dynamically attached to restrict some features based on a client's authorization (\Snospace~\ref{sec:background}). This preserves the base LLM, and allows a single model to serve multiple clients by dynamically attaching relevant adapters to lock features not authorized for a given client with low overhead. This makes it scale across multiple features \ref{scalable} $\rightarrow$ \colorbox{green!8}{\cmark}). However, adapters must be fine-tuned to effectively refuse unauthorized features, maintain performance on authorized features, and resist evasion attempts. We discuss our design choices to achieve this, and present an overview of \method's design in Figure~\ref{fig:overview}.

\noindent\textbf{\underline{Training Objective.}}
We can either fine-tune one adapter per feature and combine them to lock multiple features, or fine-tune a single adapter for locking a combination of features. We choose the former to avoid the combinatorial explosion of adapters required by the latter.
To lock a feature $f$, we obtain the adapters by fine-tuning some layers of a base LLM $\pi_{\theta}$ parameterized by $\theta$ on a feature dataset $D_{f} = \{(x_i, y_i)\}$.
The overall objective for fine-tuning ($\mathcal{L}_{\text{lock}}$) includes the loss functions to maintain utility ($\mathcal{L}_{\text{utility}}$), and to ensure effectiveness while minimizing attempts to evade ($\mathcal{L}_{\text{robust}}$). We have: $\mathcal{L}_{\text{lock}} = \mathcal{L}_{\text{utility}} + \mathcal{L}_{\text{robust}}$.

\noindent\textbf{\underline{Utility-Preserving (\ref{utility}).}} We preserve the utility of $\pi_{\theta}$ during fine-tuning by computing the KL divergence with respect to its frozen reference $\pi_{\theta'}$:
% \begin{footnotesize}
\begin{equation*}
\mathcal{L}_{\text{utility}} = \mathbb{E}_{(x, y) \in D_{\text{auth}}}\bigg[\text{KL}[\pi_{\theta}(y | x) || \pi_{\theta'}(y|x)]\bigg]
\end{equation*}
% \end{footnotesize}
\noindent Here, $D_{\text{auth}}$ contains generic question and helpful (authorized) responses, unrelated to any of the locked features $f \in \mathcal{F}$ (e.g., text from Wikipedia)~\cite{ding-etal-2023-enhancing}. This is a common technique used in unlearning \cite{gao2025on} to retain the LLM's utility on basic tasks (e.g., text completion) during fine-tuning.

\noindent\textbf{\underline{Effective (\ref{effective}) and Robust (\ref{robust}).}} Prior work has shown that adding perturbations to LLM activations can reinforce alignment~\cite{quada}. Inspired by this, we ensure effective refusal of unauthorized features and robustness against evasion by augmenting refusal training with \emph{Latent Adversarial Training} (LAT)~\cite{sheshadri2025latent}.
For this, we construct a preference dataset $D_{\text{unauth}} = \{(x_i,c_i,r_i)\}$ for locking a feature $f$, using a publicly available feature dataset $D_{f}$. Each prompt $x_i$ is from $D_{f}$, which is paired with a fixed refusal response $c_i$ (as a positive sample), along with a useful response $r_i$ (as a negative sample). We use the ground truth responses $y_i$ from $D_{f}$ as $r_i$, and $c_i$ is set to be "\texttt{Sorry, you are not authorized to use the capabilities needed to solve this problem}".

\noindent\emph{Computing Sample-wise Perturbations:} First, we find the worst-case perturbation $\delta_i$ which is added to the latent activations of a set of target layers. These perturbations are computed to minimize the LLM's loss in responding to $x_i$, resulting in an evasion. We define the loss as:
\begin{equation*}
\begin{split}
    \mathcal{L}_{\text{evade}}(c_i, r_i; \delta) =\overbrace{-\log{\pi_{\theta}(c_i | \alpha(x_i, \delta))}}^{\text{Move towards $c_i$}}\\
    + \overbrace{- \log{(1 - \pi_{\theta}(r_i | \alpha(x_i, \delta))})}^{\text{Move away from $r_i$}}
\end{split}
\end{equation*}
$\alpha(x_i, \delta)$ is a function that adds $\delta$ to the LLM's latent activations for an input $x_i$, and $\epsilon$ is the perturbation budget where $||\delta||_2 \leq \epsilon$. To find the perturbations, we compute $\delta_i = \underset{\delta}{\text{argmin }} \mathcal{L}_{\text{evade}}(r_i, c_i; \delta)$.

\noindent\emph{Robust Fine-tuning for Effectiveness:} With the perturbations for different samples, we now update $\theta$ to minimize $\mathcal{L}_{\text{robust}}$, the average of $\mathcal{L}_{\text{evade}}(c_i, r_i; \delta_i)$ over all samples in $D_{\text{unauth}}$, which encourages the LLM to:
\begin{enumerate*}[label={(\roman*)}]
\item increase the likelihood of the \emph{preferred refusal completions} $c_i$,
\item decrease the probability of producing the \emph{actual correct responses} $r_i$
\end{enumerate*}. In this way, we get an adapter to robustly lock $f$.

\noindent\textbf{\underline{Merging Adapters.}}
Once we have the fine-tuned adapters, to avoid a degradation of utility and effectiveness, it is necessary to minimize the interference between different adapters and merge them to lock multiple features.
Formally, if the client is authorized to use $f_j$, we attach adapters $\{a_k:\, k\neq j\}$ to the base LLM for locking unauthorized features $f_{-j}$ (Figure \ref{fig:overview}).
During evaluation (\Snospace~\ref{sec:mergingEval}), we observe that after applying LAT, the LLM refuses unlocked features (over-refusal), as the adapters reinforce the refusal directions. This results in the LLM generating "\texttt{Sorry, sorry, ...}" for every prompt.
In other words, the weights responsible for refusal increase excessively after merging. This causes utility degradation. Consequently, existing merging methods (\S\ref{sec:background}) result in over-refusal, and a new approach is needed to address this problem.

\noindent\emph{\method Merging:} Merging multiple low-rank adapters \cite{ortiz-jimenez2023task, hu2022lora} often results in destructive weight interference \cite{gargiulo2025task}, which reinforces the weights responsible for refusal. The maximum singular value (i.e. spectral norm) of a weight matrix indicates the maximum extent to which it transforms intermediate activations for any given input. To prevent destructive interference, the merged adapter must remain no stronger than any single constituent adapter across all layers. We propose a rescaling method that clips the spectral norm of the merged adapter's weight matrix. Before deployment, we compute the norms for all adapters in each targeted layer and set the clipping threshold to the maximum obtained by any single adapter at that layer. Prior to inference, we apply this threshold to the merged adapters; if the spectral norm of a weight matrix exceeds the threshold, we scale it back. Full approach summarized in Algorithm~\ref{algo:locket}.

\begin{algorithm}[!ht]
\small
\caption{\method \textsc{Merging}}
\textbf{Global:} $F=\{f_1,\dots,f_m\}$; adapters $\{\Delta W_\ell^{i}\}$; base weights $\{W_\ell\}$; scale $\tau$;
% black-box \textsc{GetFeatures}$(\mathrm{ctx})\!\to\! S\subseteq F$.
\\
\textbf{Offline (once):} \textsc{ComputeClippingThresholds}
\begin{algorithmic}[1]
    \For{each layer $\ell$}
    \State $Clip_\ell \leftarrow \tau \cdot \max_{i}\|\Delta W_\ell^{i}\|_2$ \Comment{per-layer threshold}
    \EndFor
\end{algorithmic}

\textbf{Online (for each client session):} \textsc{MergeAdapters}
\begin{algorithmic}[1]
    \For{each layer $\ell$}
    \State $\Delta W_\ell \leftarrow \sum_{i\in L} \Delta W_\ell^{i}$ \Comment{CAT merge}
    \If{$\|\Delta W_\ell\|_2 > Clip_\ell$}
    \State $\Delta W_\ell \leftarrow \frac{Clip_\ell}{\|\Delta W_\ell\|_2}\,\Delta W_\ell$ \Comment{post-merge rescale}
    \EndIf
    \State $W_\ell^{'} \leftarrow W_\ell + \Delta W_\ell$ \Comment{attach}
    \EndFor
\end{algorithmic}

\textbf{Online (for each inference request):} \textsc{Inference}$(x)$
\begin{algorithmic}[1]
    \State \Return $\pi_{\theta(W^{'})}(x)$ \Comment{inference response}
\end{algorithmic}
\label{algo:locket}
\end{algorithm}

\noindent Formally, during the offline stage, for each layer $\ell$ where we attach adapter update matrices, we first apply \emph{singular value decomposition} \cite{demmel1997applied} to decompose the update matrices $\Delta W_{\ell}^{i}$ of adapters $a_i$ as $\Delta W_{\ell}^{i} \approx \mathbf{U}^i\mathbf{S}^i(\mathbf{V}^i)^T$. Here, $\mathbf{S}^i = \text{diag}(\sigma_1^i, \sigma_2^i \cdots, \sigma_r^i)$ is a diagonal matrix of singular values with $\sigma_1^i \geq \sigma_2^i \geq \cdots \sigma_r^i$, and $\mathbf{U}^i, \mathbf{V}^i$ are the left and right singular vector matrices, respectively.
Then the largest singular value of each adapter matrix is $\sigma^i := \sigma_1^i = ||\Delta W_{\ell}^i||_2$, we compute a reference scale for the layer: $\sigma_{\ell} = \max{(\sigma^1, \sigma^2 \cdots, \sigma^m)}$. Then, we set the maximum norm value for the weights as $Clip_{\ell} := \tau \sigma_{\ell}$, where $0 < \tau \leq 1$ is an adjustable scaling hyperparameter.

During the online stage, we first merge the LoRA adapters via CAT. Then for each layer $\ell$, we compute the spectral norm of the merged weight matrix $\Delta W_{\ell}$. If it is greater than $Clip_{\ell}$, then we rescale it as $\Delta W_{\ell} \leftarrow f_{\ell}\Delta W_{\ell}$ where $f_{\ell} = \frac{Clip_{\ell}}{\sigma^i}$.
In this way, \method merging preserves the unlocked utility, the prominence of the refusal directions, while reducing the influence of over-refusal weights after merging. A comparison between \method merging and other approaches can be found in \Snospace~\ref{sec:mergingEval}.

% \begin{takeaway}
%     \textbf{Takeaway:} \method's design explicitly accounts for the requirements (\ref{effective}-\ref{scalable}).
% \end{takeaway}

%% file: 5setup.tex
\section{Experimental Setup}\label{sec:setup}

\noindent\textbf{Datasets.} We use four datasets, each corresponding to a premium feature:
\begin{enumerate*}[label={(\roman*)}]
\item \emph{Math (\competitionmath)} which contains challenging math problems~\cite{hendrycks2021measuringmath};
\item \emph{SQL (\sql)} for structured query language generation~\cite{zhongSeq2SQL2017, yu2018spider};
\item \emph{Text Summarization (\samsum)} from the SAMSum dataset~\cite{gliwa2019samsum};
\item \emph{General Knowledge (\mmlu)} from the MMLU benchmark~\cite{hendrycks2021measuring}.
\end{enumerate*}
% , a diverse collection of multiple-choice questions covering 57 subjects, to represent a general-purpose reasoning feature.
For $D_{\text{auth}}$, we use samples from the UltraChat dataset~\cite{ding-etal-2023-enhancing}.
We describe details about the datasets in Appendix~\ref{app:implementation}.

\noindent\textbf{Models.} Following \citet{greenblatt2024alignment}, we use two LLMs: \deepseekmath (specialized in math)~\cite{shao2024deepseekmathpushinglimitsmathematical}, and \deepseekcoder (specialized in coding)~\cite{guo2024deepseekcoderlargelanguagemodel}. We additionally use one general-purpose conversation model, \llama~\cite{grattafiori2024llama3herdmodels}.
\deepseekcoder, trained on code, includes a system prompt that refuses non-coding queries. Unlike other models, we evaluate \deepseekcoder on coding tasks.
We describe fine-tuning hyperparameters in Appendix~\ref{app:implementation}. Evaluations are conducted with temperature set to zero.

\noindent\textbf{Metrics.}
We evaluate \method based on the four key requirements defined in \Snospace~\ref{sec:problem}.
\begin{itemize}[leftmargin=*]
        \setlength{\itemsep}{0pt}
        \setlength{\parskip}{0pt}
        \setlength{\parsep}{0pt}
    \item \noindent\emph{\ref{effective} Effectiveness:} We measure utility (see below) on the test set of the locked feature, with $0.00$ indicating effective locking.
    \item \emph{\ref{utility} Utility:} We use different metrics depending on the task:
        \begin{enumerate*}[label={(\roman*)}]
        \item for \emph{Math \& MMLU}, we use accuracy wrt. ground truth answers; and
        \item for \emph{SQL \& Summarization}, we use the Rouge-1 score \cite{lin2004rouge} to evaluate the quality of generated outputs (also referred to ``accuracy'').
        \end{enumerate*}
    \item \noindent\emph{\ref{robust} Robustness:} We use attack success rate (ASR) based on how often the LLM generates responses without refusal keywords like \texttt{``sorry"}, \texttt{``I cannot"}, or \texttt{``unable"}.
    \item \noindent\emph{\ref{scalable} Scalability:} We evaluate using metrics for \ref{effective} and \ref{utility}, but after locking multiple features (e.g., \competitionmath+ \sql, \competitionmath+ \sql+ \samsum).
\end{itemize}

\noindent\textbf{Comparison with Prior Work.}
We use the password-locking technique (referred as ``PWD'') proposed by \citet{greenblatt2024stress} and used in related works (\S\ref{sec:background}), where prompts with correct password produce a useful response. Instead of locking by generating a useless response~\cite{greenblatt2024stress}, we fine-tune for refusal which resembles \citet{tangsecure}.

%% file: 6results.tex
\section{Evaluation}\label{sec:evaluation}

\input{figures/fig_merging_tau_trend}

\subsection{Evaluation of \method Merging}\label{sec:mergingEval}

We compare \method merging with existing merging methods in terms of their impact on utility-preservation. We unlock one feature and lock all remaining (to capture the worst-case impact) by merging and attaching the corresponding adapters to the base LLM.
We then measure the unlocked feature's refusal rate ($100-\text{utility}$) to determine if the merging causes over-refusal on the unlocked feature.
Figure~\ref{fig:merging} (Top) shows that \method yields significantly lower refusal rates than the others \change{(detailed results in Appendix \ref{app:merging_techniques})}

\begin{submit}
    \input{tables/tab_effectiveness}
    \input{tables/tab_utility}
\end{submit}

\noindent\textbf{Selecting the Scaling Hyperparameter $\tau$.} To show the trade-offs from varying $\tau$, we illustrate using \deepseekmath by locking three features (\competitionmath, \sql, and \samsum), and leaving \mmlu as unlocked (Figure~\ref{fig:merging}: Bottom). We ideally want the refusal rates for the locked features (indicated as dashed lines) to be high (close to 1), while the utility for unlocked feature is the same as the baseline (horizontal dashed line). We find that the value of $\tau = 0.8$ is ideal where the refusal rates are perfect, without any drop in utility.
We use the same hyperparameter tuning approach for selecting $\tau$, to lock other features and their combinations (Appendix~\ref{app:implementation}).

\begin{submit}
    \subsection{\ref{effective} (Effectiveness)}\label{sec:evalEffective}

    \noindent\textbf{Results for \method.} We evaluate effectiveness of feature locking by measuring the utility wrt. the locked feature. Ideally, this is zero (indicating 100\% refusal rate).
    Table~\ref{tab:effect_utility} (left) shows the results, where effective locking is in \colorbox{blue!10}{blue}, and ineffective locking in \colorbox{orange!20}{orange}. Columns represent locked features, while each row indicates a feature whose utility is measured. For perfect effectiveness, the utility should be zero in cells where the rows match the locked features in the columns.
    We observe this across features and models (i.e., a $100\%$ refusal rate for queries on locked features), demonstrating \method's effectiveness. % in locking unauthorized features.

    \noindent\textbf{Comparison with PWD.}
    Table~\ref{tab:scalability_vs_baselines} (left) is a comparison of \method and PWD, using \deepseekmath with ``Math'' (\competitionmath) locked.
    We expect that in cells where the \competitionmath column is locked, the utility of the \emph{same} features along the rows should be zero (\colorbox{blue!10}{blue}). This is the case for both PWD and \method, indicating perfect effectiveness.

    \subsection{\ref{utility} (Utility-Preserving)}\label{sec:evalUtility}

    \noindent\textbf{Results for \method.} Table~\ref{tab:effect_utility} (left) shows the utility of \method. Columns represent locked features, while each row indicates a feature whose utility is measured. For perfect utility, cells in rows that do not match the locked features in the columns (with non-zero values in the ``baseline'' column), should correspond to the values in the ``baseline'' column.
    We use \colorbox{green!16}{green} to indicate outperforming/matching baseline, \colorbox{yellow!20}{yellow} within $\pm$5\% of baseline, and \colorbox{red!20}{red} worse than baseline.
    When locking single features, \method can successfully preserve the utility of unlocked features in two of the three rows (\colorbox{green!16}{green}).
    Locking \competitionmath results in a small utility drop in \mmlu (\colorbox{yellow!20}{yellow}), and locking \mmlu causes a drop in \competitionmath (\colorbox{red!20}{red}). This is due to overlapping math-related questions in \mmlu, such interference among features results in a utility drop (discussed in \Snospace~\ref{sec:discussions}).

    \noindent\textbf{Comparison with PWD.}
    Table~\ref{tab:scalability_vs_baselines} (left) is a comparison of \method and PWD, using \deepseekmath with ``Math'' (\competitionmath) locked.
    For rows corresponding to \sql, \samsum, and \mmlu, we find that \method's utility matches the baseline for \sql and \samsum, but incurs a small drop for \mmlu (due to interference with \competitionmath).
    On the other hand, PWD incurs a significant $12\%$ drop in \samsum's utility (\colorbox{red!10}{red}), and a small drop for \mmlu (\colorbox{yellow!20}{yellow}).
    These suggest that \method maintains utility better than PWD.
    This likely occurs because PWD fine-tuning overwrites the base LLM’s weights, while \method modifies specific layers of the frozen base LLM, preserving utility.

    \subsection{\ref{robust} (Robustness)}\label{sec:evalRob}

    To measure the upper bound on robustness, we evaluate using \optJBplus (instead of \optJB), and consider the following attacks: \emph{Many-shot}~\cite{anil2024manyshot}, \emph{GCG}~\cite{zou2023gcg}, and \emph{AutoDAN-Turbo}~\cite{liu2025autodanturbo} (hyperparameters in Appendix~\ref{app:implementation}).
    We also evaluated against \manualJB, and found them to be ineffective against both the PWD and \method. So, we omit those results.
    Table~\ref{tab:robustness} shows the ASR for all attacks in the format ``<PWD> | \method''.
    In all cases, \method is more robust than PWD for all attacks due to latent adversarial training (\colorbox{green!16}{green}).
    Unlike prior work~\cite{greenblatt2024alignment,tangsecure}, \method uses no secret credentials, and is protected against credential stealing and redistribution (\ref{keysteal}).
    \input{tables/tab_robustness}

    \subsection{\ref{scalable} (Scalable)}\label{sec:evalScal}

    To demonstrate scalability of \method and compare with PWD, we evaluate effectiveness and utility on locking multiple features.
    Table~\ref{tab:effect_utility} (right), where the columns have two or more features, shows the results for \method.
    Across all cases, we see that \method is perfectly effective (\colorbox{blue!10}{blue}) with some drop in utility due to interference between \competitionmath and \mmlu (\colorbox{yellow!20}{yellow}).
    Table~\ref{tab:scalability_vs_baselines} (right), with two or three features locked (\competitionmath + \sql, \competitionmath + \sql + \samsum), is a comparison of \method and PWD.
    In most cases, utility and effectiveness of PWD are worse than \method  (\colorbox{orange!10}{orange}, \colorbox{yellow!20}{yellow} or \colorbox{red!10}{red}).
    This suggests that \method scales to more than two features, unlike PWD. PWD's full fine-tuning likely causes ``catastrophic forgetting''~\cite{kotha2024understanding}, where training refusal for one feature harms others.

    \begin{takeaway}
        \textbf{Takeaway:} \method outperforms PWD and meets all requirements (\ref{effective}-\ref{scalable}).
    \end{takeaway}
\end{submit}

%% file: figures/fig_merging_tau_trend.tex
\begin{figure}[!t]
\centering
\resizebox{.9\columnwidth}{!}{
  \includegraphics[width=\columnwidth]{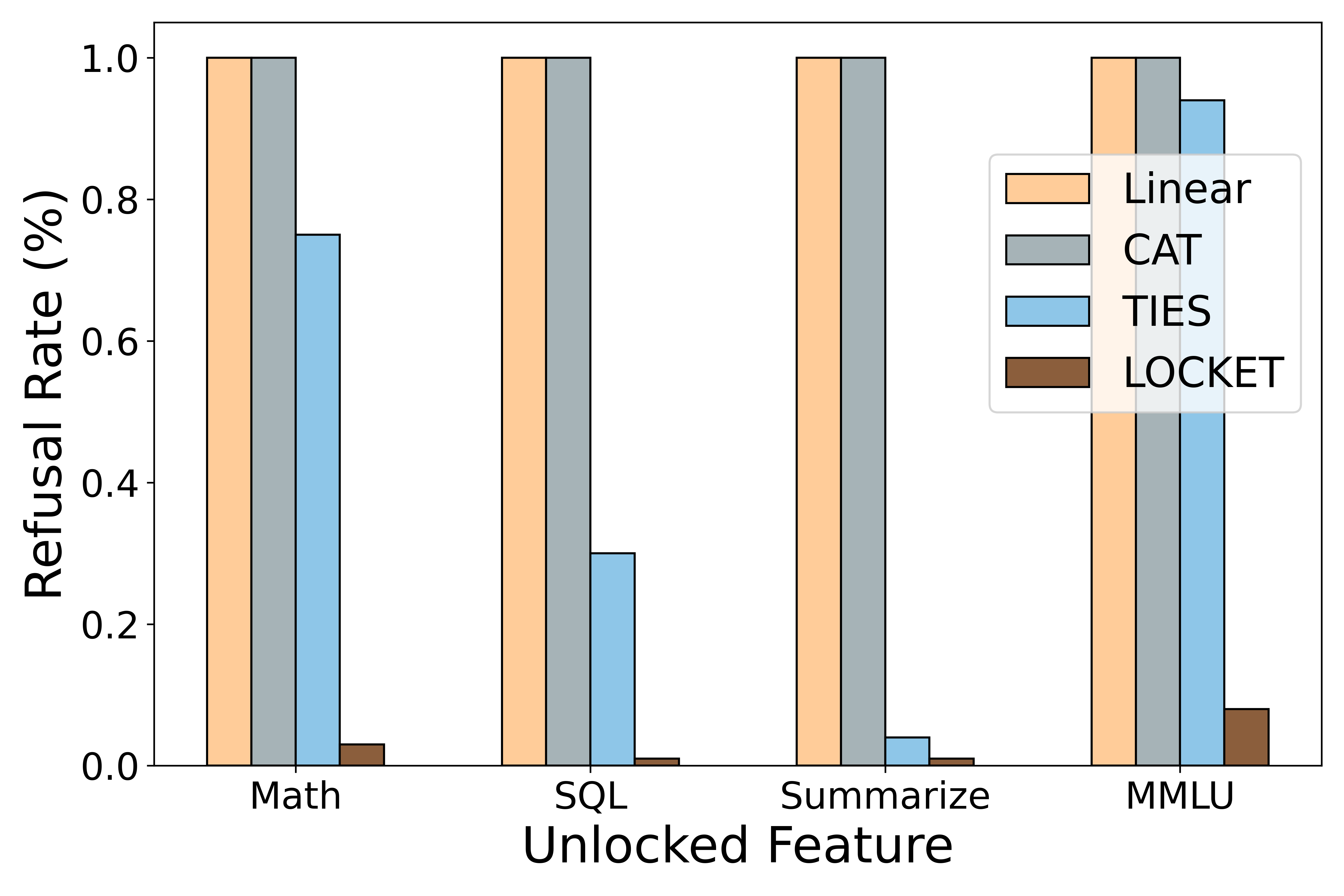}
}\\
\resizebox{.9\columnwidth}{!}{
  \includegraphics[width=\columnwidth]{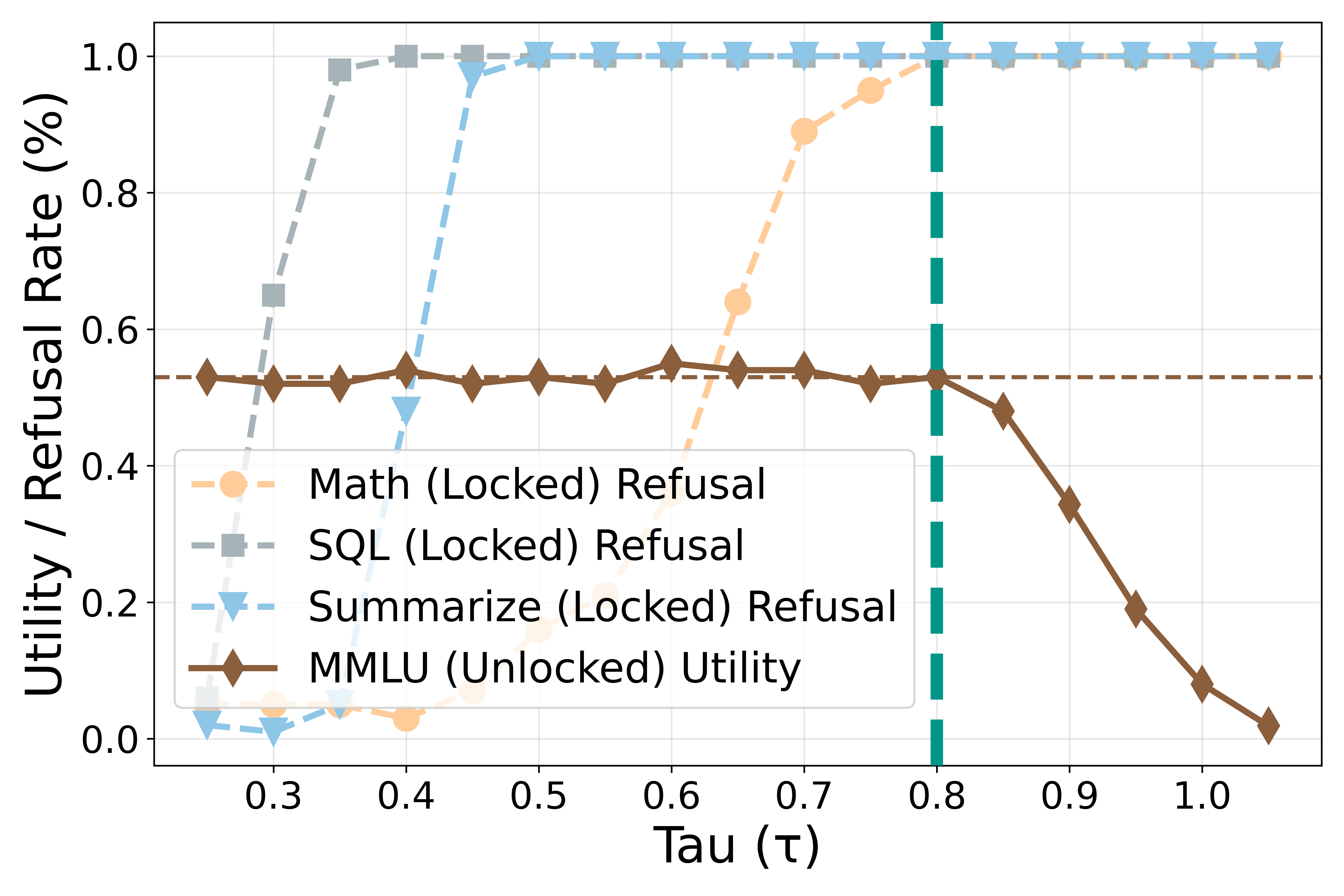}
}
  \caption{\textbf{Illustrative examples for \deepseekmath:} We observe similar patterns in other models and locking combinations. \textbf{(Top)} \method merging significantly reduces \mmlu's over-refusal rate compared to prior merging methods. \textbf{(Bottom)} The scaling hyperparameter $\tau$ should be chosen to balance the trade-off between effectiveness and utility. Here, only \mmlu is unlocked; the vertical line indicates the sweet spot for $\tau$ (high refusal for locked and no utility drop on unlocked features). See Appendix~\ref{app:implementation} for other $\tau$ values.}
  \label{fig:merging}
\end{figure}

%% file: tables/tab_effectiveness.tex
\begin{table*}[!ht]
\caption{\textbf{\method is effective, utility-preserving, and scalable}: ``Baseline'' is the model without \flote. For effectiveness (\ref{effective}), we use \colorbox{blue!10}{blue} to indicate complete locking and \colorbox{orange!20}{orange} otherwise. For utility (\ref{utility}) (of unlocked features) \colorbox{green!16}{green} $\rightarrow$ matches/outperforms baseline, \colorbox{yellow!20}{yellow} $\rightarrow$ within $\pm$5\% of baseline, \colorbox{red!20}{red} $\rightarrow$ worse than baseline. Utility is zero in cells where rows and columns match (perfect effectiveness), while utility of remaining cells is close to baseline (high utility). We see this even when multiple features are locked, demonstrating scalability.}
\centering
\renewcommand{\rothead}[1]{\rotatebox[origin=t]{90}{#1}}
\begingroup
\footnotesize
\setlength{\tabcolsep}{3pt}
\def\arraystretch{1.2}
\resizebox{\textwidth}{!}{
\begin{scriptsize}
\begin{tabularx}{\textwidth}{l|*{5}{C}|*{11}{C}}
\bottomrule

\toprule

\multicolumn{1}{l|}{\textbf{Locked Feature $\rightarrow$}} &
\rothead{\textbf{Baseline}} &
\rothead{\textbf{Math} (\competitionmath)} &
\rothead{\textbf{SQL} (\sql)} &
\rothead{\textbf{Summarize} (\samsum)} &
\rothead{\textbf{MMLU} (\mmlu)} &
\rothead{\competitionmath + \sql} &
\rothead{\competitionmath + \samsum} &
\rothead{\competitionmath + \mmlu} &
\rothead{\sql + \samsum} &
\rothead{\sql + \mmlu} &
\rothead{\samsum + \mmlu} &
\rothead{\competitionmath + \sql + \samsum} &
\rothead{\competitionmath + \sql + \mmlu} &
\rothead{\competitionmath + \samsum + \mmlu} &
\rothead{\sql + \samsum + \mmlu} &
\rothead{~~\competitionmath + \sql + \samsum + \mmlu~~} \\

\multicolumn{1}{l|}{\textbf{Evaluated Feature $\downarrow$}} & \multicolumn{5}{c|}{For Effectiveness (\ref{effective}) and Utility (\ref{utility})} & \multicolumn{11}{c}{Scalability (\ref{scalable}) w.r.t Effectiveness (\ref{effective}) and Utility (\ref{utility})} \\

\midrule

\multicolumn{17}{c}{\rule{0pt}{2ex}\textbf{\deepseekmath locked via \method}} \\

\midrule

\textbf{Math} (\competitionmath) & 0.40 & \diffcell{0.40}{0.00}0.00 & \diffcell{0.40}{0.45}0.45 & \diffcell{0.40}{0.40}0.40 & \diffcell{0.40}{0.42}0.42 & \diffcell{0.40}{0.00}0.00 & \diffcell{0.40}{0.00}0.00 & \diffcell{0.40}{0.00}0.00 & \diffcell{0.40}{0.43}0.43 & \diffcell{0.40}{0.44}0.44 & \diffcell{0.40}{0.44}0.44 & \diffcell{0.40}{0.00}0.00 & \diffcell{0.40}{0.00}0.00 & \diffcell{0.40}{0.00}0.00 & \diffcell{0.40}{0.45}0.45 & \diffcell{0.62}{0.00}0.00 \\

\textbf{SQL} (\sql) & 0.93 & \diffcell{0.93}{0.95}0.95 & \diffcell{0.93}{0.00}0.00 & \diffcell{0.93}{0.93}0.93 & \diffcell{0.93}{0.93}0.93 & \diffcell{0.93}{0.00}0.00 & \diffcell{0.93}{0.94}0.94 & \diffcell{0.93}{0.94}0.94 & \diffcell{0.93}{0.00}0.00 & \diffcell{0.93}{0.00}0.00 & \diffcell{0.93}{0.93}0.93 & \diffcell{0.93}{0.00}0.00 & \diffcell{0.93}{0.00}0.00 & \diffcell{0.93}{0.94}0.94 & \diffcell{0.93}{0.00}0.00 & \diffcell{0.95}{0.00}0.00 \\

\textbf{Summarize} (\samsum) & 0.23 & \diffcell{0.23}{0.23}0.23 & \diffcell{0.23}{0.24}0.24 & \diffcell{0.23}{0.00}0.00 & \diffcell{0.23}{0.24}0.24 & \diffcell{0.23}{0.24}0.24 & \diffcell{0.23}{0.00}0.00 & \diffcell{0.23}{0.24}0.24 & \diffcell{0.23}{0.00}0.00 & \diffcell{0.23}{0.24}0.24 & \diffcell{0.23}{0.00}0.00 & \diffcell{0.23}{0.00}0.00 & \diffcell{0.23}{0.24}0.24 & \diffcell{0.23}{0.00}0.00 & \diffcell{0.23}{0.00}0.00 & \diffcell{0.27}{0.00}0.00 \\

\textbf{MMLU} (\mmlu) & 0.53 & \cellcolor{yellow!20}0.51 & \cellcolor{yellow!20}0.50 & \diffcell{0.53}{0.53}0.53 & \diffcell{0.53}{0.00}0.00 & \diffcell{0.53}{0.53}0.53 & \diffcell{0.53}{0.53}0.53 & \diffcell{0.53}{0.00}0.00 & \diffcell{0.53}{0.54}0.54 & \diffcell{0.53}{0.00}0.00 & \diffcell{0.53}{0.00}0.00 & \diffcell{0.53}{0.53}0.53 & \diffcell{0.53}{0.00}0.00 & \diffcell{0.53}{0.00}0.00 & \diffcell{0.53}{0.00}0.00 & \diffcell{0.53}{0.00}0.00 \\

\midrule

\multicolumn{17}{c}{\rule{0pt}{1ex}\textbf{\deepseekcoder locked via \method}} \\

\midrule

\textbf{SQL} (\sql) & 0.96 & \diffcell{0.96}{0.96}0.96 & \diffcell{0.96}{0.00}0.00 & \diffcell{0.96}{0.96}0.96 & \diffcell{0.96}{0.96}0.96 & \diffcell{0.96}{0.00}0.00 & \cellcolor{yellow!20}0.93 & \diffcell{0.96}{0.96}0.96 & \diffcell{0.96}{0.00}0.00 & \diffcell{0.96}{0.00}0.00 & \cellcolor{yellow!20}0.95 & \diffcell{0.96}{0.00}0.00 & \diffcell{0.96}{0.00}0.00 & \diffcell{0.96}{0.96}0.96 & \diffcell{0.96}{0.00}0.00 & \diffcell{0.96}{0.00}0.00 \\

\midrule

\multicolumn{17}{c}{\rule{0pt}{2ex}\textbf{\llama locked via \method}} \\

\midrule

\textbf{Math} (\competitionmath) & 0.28 & \diffcell{0.28}{0.00}0.00 & \diffcell{0.28}{0.28}0.28 & \diffcell{0.28}{0.28}0.28 & \cellcolor{red!20}0.22 & \diffcell{0.21}{0.00}0.00 & \diffcell{0.21}{0.00}0.00 & \diffcell{0.21}{0.00}0.00 & \cellcolor{yellow!20}0.27 & \cellcolor{red!20}0.21 & \cellcolor{yellow!20}0.23 & \diffcell{0.21}{0.00}0.00 & \diffcell{0.21}{0.00}0.00 & \diffcell{0.21}{0.00}0.00 & \diffcell{0.88}{0.00}0.00 & \cellcolor{yellow!20}0.23 \\

\textbf{SQL} (\sql) & 0.88 & \diffcell{0.88}{0.92}0.92 & \diffcell{0.88}{0.00}0.00 & \diffcell{0.88}{0.93}0.93 & \diffcell{0.88}{0.89}0.89 & \diffcell{0.88}{0.00}0.00 & \diffcell{0.88}{0.93}0.93 & \diffcell{0.88}{0.92}0.92 & \diffcell{0.88}{0.00}0.00 & \diffcell{0.88}{0.00}0.00 & \diffcell{0.88}{0.92}0.92 & \diffcell{0.88}{0.00}0.00 & \diffcell{0.88}{0.00}0.00 & \diffcell{0.88}{0.89}0.89 & \diffcell{0.88}{0.00}0.00 & \diffcell{0.88}{0.00}0.00 \\

\textbf{Summarize} (\samsum) & 0.32 & \diffcell{0.32}{0.34}0.34 & \diffcell{0.32}{0.32}0.32 & \diffcell{0.32}{0.00}0.00 & \diffcell{0.32}{0.32}0.32 & \diffcell{0.32}{0.34}0.34 & \diffcell{0.32}{0.00}0.00 & \diffcell{0.32}{0.33}0.33 & \diffcell{0.32}{0.00}0.00 & \diffcell{0.32}{0.32}0.32 & \diffcell{0.32}{0.00}0.00 & \diffcell{0.32}{0.00}0.00 & \diffcell{0.32}{0.33}0.33 & \diffcell{0.32}{0.00}0.00 & \diffcell{0.32}{0.00}0.00 & \diffcell{0.32}{0.00}0.00 \\

\textbf{MMLU} (\mmlu)  & 0.67 & \cellcolor{yellow!20}0.64 & \diffcell{0.67}{0.71}0.71 & \diffcell{0.67}{0.68}0.68 & \diffcell{0.63}{0.00}0.00 &  \diffcell{0.67}{0.73}0.73 & \diffcell{0.67}{0.70}0.70 & \diffcell{0.67}{0.00}0.00 & \diffcell{0.67}{0.69}0.69 & \diffcell{0.63}{0.00}0.00 & \diffcell{0.63}{0.00}0.00 & \diffcell{0.67}{0.72}0.72 & \diffcell{0.63}{0.00}0.00 & \diffcell{0.63}{0.00}0.00 & \diffcell{0.63}{0.00}0.00 & \diffcell{0.63}{0.00}0.00 \\

\bottomrule

\toprule
\end{tabularx}
\end{scriptsize}
}
\endgroup
\label{tab:effect_utility}
\end{table*}

%% file: tables/tab_utility.tex
\begin{table*}[!ht]
\caption{\textbf{Comparison of \method with prior work:} We compare effectiveness and utility of \method with prior work (``PWD'')~\cite{greenblatt2024alignment,tangsecure} locking \deepseekmath. ``Baseline'' is the model without \flote. 
Color coding for effectiveness, utility, and scalability, are same as Table~\ref{tab:effect_utility}.}
% For effectiveness, we look at the rows where baseline is zero, effective locking is \colorbox{blue!10}{blue} and \colorbox{orange!20}{orange} otherwise. Rest of the baseline values are to evaluate utility, where \colorbox{green!16}{green} $\rightarrow$ matches baseline, \colorbox{yellow!20}{orange} $\rightarrow$ within $\pm$5\% of baseline, \colorbox{red!20}{red} $\rightarrow$ worse than baseline.}
\centering
\begingroup
\footnotesize
\setlength{\tabcolsep}{3pt}
\def\arraystretch{1.2}
\resizebox{\textwidth}{!}{
\begin{scriptsize}
\begin{tabularx}{\textwidth}{l|CCC|CCCCCC}
\bottomrule

    \toprule

    \textbf{Locked Feature $\rightarrow$} & \multicolumn{3}{c|}{\competitionmath (For \ref{effective}: Effectiveness; \ref{utility}: Utility)} & \multicolumn{3}{c}{\competitionmath + \sql (\ref{scalable}: Scalability w.r.t. \ref{effective} and \ref{utility})} & \multicolumn{3}{c}{\competitionmath + \sql + \samsum (\ref{scalable}: Scalability w.r.t. \ref{effective} and \ref{utility})}\\

    \cmidrule(lr){2-4} \cmidrule(lr){5-7} \cmidrule(lr){8-10}

    \textbf{Eval. Feature $\downarrow$} & \textbf{Baseline} & \textbf{PWD} & \textbf{\method} & \textbf{Baseline} & \textbf{PWD} & \textbf{\method} & \textbf{Baseline} & \textbf{PWD} & \textbf{\method} \\
    
    \midrule

    % \textbf{Math} (\competitionmath) & 0.00 & \diffcell{0.00}{0.00}0.00 & \diffcell{0.00}{0.00}0.00 & 0.00 & \nediffcell{0.00}{0.35}0.35 & \diffcell{0.00}{0.00}0.00 & 0.00 & \nediffcell{0.00}{0.26}0.26 & \diffcell{0.00}{0.00}0.00 \\

    % \textbf{Code} (\sql) & 0.93 & \diffcell{0.93}{0.93}0.93 & \diffcell{0.93}{0.95}0.95 & 0.00 & \diffcell{0.00}{0.00}0.00 & \diffcell{0.00}{0.00}0.00 & 0.00 & \diffcell{0.00}{0.00}0.00 & \diffcell{0.00}{0.00}0.00 \\
    
    % \textbf{Summarize} (\samsum) & 0.23 & \nediffcell{0.23}{0.11}0.11 & \diffcell{0.23}{0.23}0.23 & 0.23 & \diffcell{0.23}{0.27}0.27 & \diffcell{0.23}{0.24}0.24 & 0.00 & \nediffcell{0.00}{0.12}0.12 & \diffcell{0.00}{0.00}0.00 \\
    
    % \textbf{MMLU} (\mmlu) & 0.54 & \nediffcell{0.54}{0.51}0.51 & \nediffcell{0.54}{0.49}0.49 & 0.54 & \nediffcell{0.54}{0.50}0.50 & \nediffcell{0.54}{0.51}0.51 & 0.54 & \nediffcell{0.54}{0.46}0.46 & \nediffcell{0.54}{0.53}0.53 \\

    \textbf{Math} (\competitionmath) & 0.00 & \cellcolor{blue!10}0.00 & \cellcolor{blue!10}0.00 & 0.00 & \cellcolor{orange!20}0.35 & \cellcolor{blue!10}0.00 & 0.00 & \cellcolor{orange!20}0.26 & \cellcolor{blue!10}0.00 \\
    % \cmidrule{2-4}
    \hhline{~---~}
    
    \textbf{SQL} (\sql) & 0.93 & \diffcell{0.93}{0.93}0.93 & \diffcell{0.93}{0.95}0.95 & 0.00 & \cellcolor{blue!10}0.00 & \cellcolor{blue!10}0.00 & 0.00 & \cellcolor{blue!10}0.00 & \cellcolor{blue!10}0.00 \\
    % \cmidrule{5-8}
    \hhline{~~~~---~}
    
    \textbf{Summarize} (\samsum) & 0.23 & \cellcolor{red!20}0.11 & \cellcolor{green!16}0.23 & 0.23 & \diffcell{0.23}{0.27}0.27 & \diffcell{0.23}{0.24}0.24 & 0.00 & \cellcolor{orange!20}0.12 & \cellcolor{blue!10}0.00 \\
    % \cmidrule{5-8}
    \hhline{~~~~~~~---}
    
    \textbf{MMLU} (\mmlu) & 0.53 & \cellcolor{yellow!20}0.51 & \cellcolor{yellow!20}0.51 & 0.53 & \cellcolor{yellow!20}0.50 & \cellcolor{yellow!20}0.51 & 0.53 & \cellcolor{red!20}0.46 & \cellcolor{green!16}0.53 \\

\bottomrule

    \toprule
\end{tabularx}
\end{scriptsize}
}
\endgroup
\label{tab:scalability_vs_baselines}
\end{table*}

%% file: tables/tab_robustness.tex
\begin{table}[!hb]
\caption{\textbf{\method is more robust than PWD:} Attack success rates (ASR) on adversarial prompts (lower is better). Results are written as ``<PWD> | <\method>'' with \colorbox{green!16}{green} indicating \method outperforms PWD, \colorbox{red!20}{red} for worse, and \colorbox{yellow!20}{yellow} for similar (within $\pm$5\%).}
\centering
\begin{scriptsize}
\begin{tabularx}{\columnwidth}{c|CCCC}
    \bottomrule
    
    \toprule

        \textbf{Locked $\downarrow$} & \textbf{Many-shot} & \textbf{GCG} & \textbf{TAP}   & \textbf{AutoDAN} \\
        
        % \midrule

        % \multicolumn{7}{c}{\rule{0pt}{1ex}\llama locked via \cite{tangsecure}} \\

        % \midrule
        
        % \textbf{Summarize} (\samsum)  & 0.36 | 0.00 & 0.67 | 0.00 & 0.72 | 0.00 & 0.30 | 0.01 & 0.87 | 0.03 & 0.90 | 0.05 \\

        \midrule

        \multicolumn{5}{c}{\rule{0pt}{1ex}\textbf{\deepseekmath}} \\

        \midrule
        
        \competitionmath & \cellcolor{green!16}0.57 | 0.00 & \cellcolor{green!16}0.87 | 0.01 & \cellcolor{green!16}0.91 | 0.02 & \cellcolor{green!16}0.95 | 0.05 \\

 % \nediffcell{0.00}{0.00}0.00 & \nediffcell{0.00}{0.00}0.00 & \nediffcell{0.00}{0.01}0.01 & \nediffcell{0.00}{0.03}0.03 & \nediffcell{0.00}{0.05}0.05 \\

        \sql & \cellcolor{green!16}0.92 | 0.00 & \cellcolor{green!16}0.82 | 0.01 & \cellcolor{green!16}0.94 | 0.03 & \cellcolor{green!16}0.97 | 0.05 \\

        % \textbf{Code} (\sql) & \nediffcell{0.00}{0.00}0.00 & \nediffcell{0.00}{0.00}0.00 & \nediffcell{0.00}{0.00}0.00 & \nediffcell{0.00}{0.01}0.01 & \nediffcell{0.00}{0.05}0.05 & \nediffcell{0.00}{0.05}0.05 \\

        \samsum & \cellcolor{green!16}0.64 | 0.00 & \cellcolor{green!16}0.25 | 0.02 & \cellcolor{green!16}0.79 | 0.03 & \cellcolor{green!16}0.88 | 0.04\\

        % \textbf{Summarize} (\samsum) & \nediffcell{0.00}{0.00}0.00 & \nediffcell{0.00}{0.00}0.00 & \nediffcell{0.00}{0.00}0.00 & \nediffcell{0.00}{0.02}0.02 & \nediffcell{0.00}{0.03}0.03 & \nediffcell{0.00}{0.03}0.03 \\

        \mmlu & \cellcolor{green!16} 0.12 | 0.02 & \cellcolor{green!16} 0.65 | 0.03 & \cellcolor{green!16} 0.78 | 0.03 & \cellcolor{green!16} 0.89 | 0.04 \\

        % \textbf{MMLU} (\mmlu) & \nediffcell{0.00}{0.00}0.00 & \nediffcell{0.00}{0.00}0.00 & \nediffcell{0.00}{0.02}0.02 & \nediffcell{0.00}{0.03}0.03 & \nediffcell{0.00}{0.03}0.03 & \nediffcell{0.00}{0.05}0.05 \\

        \midrule

        \multicolumn{5}{c}{\rule{0pt}{1ex}\textbf{\deepseekcoder}} \\

        \midrule

        \sql & \cellcolor{green!16} 0.92 | 0.01 & \cellcolor{green!16} 0.54 | 0.02 & \cellcolor{green!16} 0.94 | 0.03 & \cellcolor{green!16} 0.96 | 0.05 \\
   % \textbf{Code} (\sql) & \nediffcell{0.00}{0.00}0.00 & \nediffcell{0.00}{0.00}0.00 & \nediffcell{0.00}{0.01}0.01 & \nediffcell{0.00}{0.02}0.02 & \nediffcell{0.00}{0.03}0.03 & \nediffcell{0.00}{0.05}0.05 \\
        \midrule

        \multicolumn{5}{c}{\rule{0pt}{1ex}\textbf{\llama}} \\

        \midrule
        
        \competitionmath & \cellcolor{green!16} 0.90 | 0.00 & \cellcolor{green!16} 0.28 | 0.01 & \cellcolor{green!16} 0.90 | 0.03 & \cellcolor{green!16} 0.93 | 0.05 \\

% & \nediffcell{0.00}{0.00}0.00 & \nediffcell{0.00}{0.00}0.00 & \nediffcell{0.00}{0.00}0.00 & \nediffcell{0.00}{0.01}0.01 & \nediffcell{0.00}{0.03}0.03 & \nediffcell{0.00}{0.05}0.05 \\

        \sql  & \cellcolor{green!16} 0.26 | 0.00 & \cellcolor{green!16} 0.30 | 0.02 & \cellcolor{green!16} 0.55 | 0.02 & \cellcolor{green!16} 0.69 | 0.03 \\

        % \textbf{Code} (\sql) & \nediffcell{0.00}{0.00}0.00 & \nediffcell{0.00}{0.00}0.00 & \nediffcell{0.00}{0.00}0.00 & \nediffcell{0.00}{0.02}0.02 & \nediffcell{0.00}{0.02}0.02 & \nediffcell{0.00}{0.03}0.03 \\

        % \textbf{Summarize} (\samsum) & \nediffcell{0.00}{0.42}0.42 & \nediffcell{0.00}{0.81}0.81 & \nediffcell{0.00}{0.82}0.82 & \nediffcell{0.00}{0.34}0.34 & \nediffcell{0.00}{0.91}0.91 & \nediffcell{0.00}{0.94}0.94 \\

        \samsum & \cellcolor{green!16} 0.72 | 0.00 & \cellcolor{green!16} 0.30 | 0.02 & \cellcolor{green!16} 0.87 | 0.03 & \cellcolor{green!16} 0.90 | 0.04 \\

        \mmlu & \cellcolor{green!16} 0.12 | 0.00 & \cellcolor{green!16} 0.39 | 0.02 & \cellcolor{green!16} 0.59 | 0.02 & \cellcolor{green!16} 0.68 | 0.03 \\

 % \nediffcell{0.00}{0.00}0.00 & \nediffcell{0.00}{0.00}0.00 & \nediffcell{0.00}{0.00}0.00 & \nediffcell{0.00}{0.02}0.02 & \nediffcell{0.00}{0.02}0.02 & \nediffcell{0.00}{0.03}0.03 \\

    \bottomrule
    
    \toprule
    \end{tabularx}
    \end{scriptsize}
\label{tab:robustness}
\end{table}

%% file: 6results_arxiv.tex
\begin{arxiv}
    \subsection{\ref{effective} (Effectiveness)}\label{sec:evalEffective}

    \noindent\textbf{Results for \method.} We evaluate effectiveness by measuring utility on locked features, which should ideally be zero (100\% refusal rate). Table~\ref{tab:effect_utility} shows the results, with effective locking in \colorbox{blue!10}{blue}. For perfect effectiveness, diagonal cells (same feature in row and column) should be zero. This holds across all cases, confirming \method's effectiveness.

    \input{tables/tab_effectiveness_arxiv}

    \noindent\textbf{Comparison with PWD.}
    Using \deepseekmath with ``Math'' (\competitionmath) locked, both PWD and \method achieve zero utility on \competitionmath, indicating perfect effectiveness.

    \subsection{\ref{utility} (Utility-Preserving)}\label{sec:evalUtility}

    \noindent\textbf{Results for \method.} Table~\ref{tab:effect_utility} also shows utility preservation, measured by performance on unlocked features relative to baseline. Non-diagonal cells should match baseline for perfect utility. We use \colorbox{green!16}{green} for matching/outperforming baseline, \colorbox{yellow!20}{yellow} for within $\pm$5\%, and \colorbox{red!20}{red} for worse.
    When locking single features, \method preserves utility in two of three rows (\colorbox{green!16}{green}). Locking \competitionmath (both \deepseekmath and \llama) or \sql (\deepseekmath) causes a small 2-3\% drop in \mmlu (\colorbox{yellow!20}{yellow}), while locking \mmlu in \llama causes a 6\% drop in \competitionmath (\colorbox{red!20}{red}) due to feature interference from math-related questions in \mmlu (\Snospace~\ref{sec:discussions}).

    \noindent\textbf{Comparison with PWD.}
    Using \deepseekmath with \competitionmath locked, \method matches baseline utility for \sql and \samsum, with only a 2\% drop on \mmlu (due to interference). In contrast, PWD shows a significant 12\% drop on \samsum and similar 2\% on \mmlu. \method better preserves utility because it augments specific layers of a frozen LLM, whereas PWD fine-tunes and overwrites original weights.

\input{tables/tab_robustness}
    \subsection{\ref{robust} (Robustness)}\label{sec:evalRob}
    \noindent Adversarial prompts transfer across models~\cite{zou2023gcg, mehrotra2024tree}: prompts evading a model with one locked feature (e.g., \competitionmath) also work when that feature is locked with others (e.g., \competitionmath+\samsum). To measure upper-bound robustness, we attack single-feature-locked models with \emph{Many-shot}~\cite{anil2024manyshot}, \emph{GCG}~\cite{zou2023gcg}, and \emph{AutoDAN-Turbo}~\cite{liu2025autodanturbo} (hyperparameters in Appendix~\ref{app:implementation}). \manualJB omitted as they were ineffective against both PWD and \method.

    Table~\ref{tab:robustness} shows ASRs as ``PWD | \method''. \method is consistently more robust than PWD (\colorbox{green!16}{green}) while maintaining effectiveness and utility. Since \method uses no secret credentials, it is also protected against credential stealing and redistribution (\ref{keysteal})~\cite{tangsecure}.

    \input{tables/tab_scalability_arxiv}

    \subsection{\ref{scalable} (Scalable)}\label{sec:evalScal}

    We evaluate scalability by measuring effectiveness (\ref{effective}) and utility (\ref{utility}) when locking multiple features. Table~\ref{tab:scalability} shows results across all models and feature combinations, with color coding as in Table~\ref{tab:effect_utility}. \method achieves perfect effectiveness (\colorbox{blue!10}{blue}) with utility drops $\leq$7\% from \competitionmath-\mmlu interference.

    Table~\ref{tab:scalability_vs_baselines} compares \method and PWD with two or three features locked (\competitionmath+ \sql, \competitionmath+ \sql+ \samsum) for \deepseekmath; other models show similar patterns (Appendix~\ref{app:scalability_results}). PWD's utility and effectiveness are mostly worse (\colorbox{orange!10}{orange}, \colorbox{yellow!20}{yellow}, or \colorbox{red!10}{red}), suggesting \method scales better. PWD's full fine-tuning likely causes ``catastrophic forgetting''~\cite{kotha2024understanding}, where training refusal for one feature harms others.

    \input{tables/tab_scalability_comparsion_arxiv}

    \begin{takeaway}
        \textbf{Takeaway:} \method outperforms PWD and meets all requirements (\ref{effective}-\ref{scalable}).
    \end{takeaway}
\end{arxiv}

%% file: tables/tab_effectiveness_arxiv.tex
\begin{table}[!ht]
\caption{\textbf{\method is effective and utility-preserving}: ``Baseline'' is the original model behaviour without \flote. For effectiveness (\ref{effective}), we use \colorbox{blue!10}{blue} to indicate complete locking. For utility (\ref{utility}) (of unlocked features) \colorbox{green!16}{green} $\rightarrow$ matches/outperforms baseline, \colorbox{yellow!20}{yellow} $\rightarrow$ within $\pm$5\% of baseline, \colorbox{red!20}{red} $\rightarrow$ worse than baseline. Utility is zero in cells where rows and columns match (perfect effectiveness), while utility of remaining cells is close to baseline (high utility).}
\centering
\renewcommand{\rothead}[1]{\rotatebox[origin=t]{70}{#1}}
\begingroup
\footnotesize
\setlength{\tabcolsep}{3pt}
\def\arraystretch{1.2}
\begin{scriptsize}
\begin{tabularx}{\columnwidth}{l|*{5}{C}}
\bottomrule

\toprule

\multicolumn{1}{l|}{\textbf{Locked Feature $\rightarrow$}} &
\rothead{\textbf{Baseline}} &
\rothead{\textbf{Math} (\competitionmath)} &
\rothead{\textbf{SQL} (\sql)} &
\rothead{\textbf{Summarize} (\samsum)} &
\rothead{\textbf{MMLU} (\mmlu)} \\

\midrule

\multicolumn{6}{c}{\rule{0pt}{2ex}\textbf{\deepseekmath locked via \method}} \\

\midrule

\textbf{Math} (\competitionmath) & 0.40 & \diffcell{0.40}{0.00}0.00 & \diffcell{0.40}{0.45}0.45 & \diffcell{0.40}{0.40}0.40 & \diffcell{0.40}{0.42}0.42 \\

\textbf{SQL} (\sql) & 0.93 & \diffcell{0.93}{0.95}0.95 & \diffcell{0.93}{0.00}0.00 & \diffcell{0.93}{0.93}0.93 & \diffcell{0.93}{0.93}0.93 \\

\textbf{Summarize} (\samsum) & 0.23 & \diffcell{0.23}{0.23}0.23 & \diffcell{0.23}{0.24}0.24 & \diffcell{0.23}{0.00}0.00 & \diffcell{0.23}{0.24}0.24 \\

\textbf{MMLU} (\mmlu) & 0.53 & \cellcolor{yellow!20}0.51 & \cellcolor{yellow!20}0.50 & \diffcell{0.53}{0.53}0.53 & \diffcell{0.53}{0.00}0.00 \\

\midrule

\multicolumn{6}{c}{\rule{0pt}{1ex}\textbf{\deepseekcoder locked via \method}} \\

\midrule

\textbf{SQL} (\sql) & 0.96 & \diffcell{0.96}{0.96}0.96 & \diffcell{0.96}{0.00}0.00 & \diffcell{0.96}{0.96}0.96 & \diffcell{0.96}{0.96}0.96 \\

\midrule

\multicolumn{6}{c}{\rule{0pt}{2ex}\textbf{\llama locked via \method}} \\

\midrule

\textbf{Math} (\competitionmath) & 0.28 & \diffcell{0.28}{0.00}0.00 & \diffcell{0.28}{0.28}0.28 & \diffcell{0.28}{0.28}0.28 & \cellcolor{red!20}0.22 \\

\textbf{SQL} (\sql) & 0.88 & \diffcell{0.88}{0.92}0.92 & \diffcell{0.88}{0.00}0.00 & \diffcell{0.88}{0.93}0.93 & \diffcell{0.88}{0.89}0.89 \\

\textbf{Summarize} (\samsum) & 0.32 & \diffcell{0.32}{0.34}0.34 & \diffcell{0.32}{0.32}0.32 & \diffcell{0.32}{0.00}0.00 & \diffcell{0.32}{0.32}0.32 \\

\textbf{MMLU} (\mmlu)  & 0.67 & \cellcolor{yellow!20}0.64 & \diffcell{0.67}{0.71}0.71 & \diffcell{0.67}{0.68}0.68 & \diffcell{0.63}{0.00}0.00 \\

\bottomrule

\toprule
\end{tabularx}
\end{scriptsize}
\endgroup
\label{tab:effect_utility}
\end{table}

%% file: tables/tab_scalability_arxiv.tex
\begin{table*}[!t]
\caption{\textbf{\method is scalable}: ``Baseline'' is the original model behaviour without \flote. For effectiveness (\ref{effective}), we use \colorbox{blue!10}{blue} to indicate complete locking. For utility (\ref{utility}) (of unlocked features) \colorbox{green!16}{green} $\rightarrow$ matches/outperforms baseline, \colorbox{yellow!20}{yellow} $\rightarrow$ within $\pm$5\% of baseline, \colorbox{red!20}{red} $\rightarrow$ worse than baseline. Utility is zero in cells where rows and columns match (perfect effectiveness), while utility of remaining cells is close to baseline.}
\centering
\renewcommand{\rothead}[1]{\rotatebox[origin=t]{70}{#1}}
\begingroup
\footnotesize
\setlength{\tabcolsep}{3pt}
\def\arraystretch{1.2}
\resizebox{\textwidth}{!}{
\begin{scriptsize}
\begin{tabularx}{\textwidth}{l|*{12}{C}}
\bottomrule

\toprule

\multicolumn{1}{l|}{\textbf{Locked Feature $\rightarrow$}} &
\rothead{\textbf{Baseline}} &
\rothead{\competitionmath + \sql} &
\rothead{\competitionmath + \samsum} &
\rothead{\competitionmath + \mmlu} &
\rothead{\sql + \samsum} &
\rothead{\sql + \mmlu} &
\rothead{\samsum + \mmlu} &
\rothead{\competitionmath + \sql + \samsum} &
\rothead{\competitionmath + \sql + \mmlu} &
\rothead{\competitionmath + \samsum + \mmlu} &
\rothead{\sql + \samsum + \mmlu} &
\rothead{~~\competitionmath + \sql + \samsum + \mmlu~~} \\

\midrule

\multicolumn{13}{c}{\rule{0pt}{2ex}\textbf{\deepseekmath locked via \method}} \\

\midrule

\textbf{Math} (\competitionmath) & 0.40 & \diffcell{0.40}{0.00}0.00 & \diffcell{0.40}{0.00}0.00 & \diffcell{0.40}{0.00}0.00 & \diffcell{0.40}{0.43}0.43 & \diffcell{0.40}{0.44}0.44 & \diffcell{0.40}{0.44}0.44 & \diffcell{0.40}{0.00}0.00 & \diffcell{0.40}{0.00}0.00 & \diffcell{0.40}{0.00}0.00 & \diffcell{0.40}{0.45}0.45 & \diffcell{0.62}{0.00}0.00 \\

\textbf{SQL} (\sql) & 0.93 & \diffcell{0.93}{0.00}0.00 & \diffcell{0.93}{0.94}0.94 & \diffcell{0.93}{0.94}0.94 & \diffcell{0.93}{0.00}0.00 & \diffcell{0.93}{0.00}0.00 & \diffcell{0.93}{0.93}0.93 & \diffcell{0.93}{0.00}0.00 & \diffcell{0.93}{0.00}0.00 & \diffcell{0.93}{0.94}0.94 & \diffcell{0.93}{0.00}0.00 & \diffcell{0.95}{0.00}0.00 \\

\textbf{Summarize} (\samsum) & 0.23 & \diffcell{0.23}{0.24}0.24 & \diffcell{0.23}{0.00}0.00 & \diffcell{0.23}{0.24}0.24 & \diffcell{0.23}{0.00}0.00 & \diffcell{0.23}{0.24}0.24 & \diffcell{0.23}{0.00}0.00 & \diffcell{0.23}{0.00}0.00 & \diffcell{0.23}{0.24}0.24 & \diffcell{0.23}{0.00}0.00 & \diffcell{0.23}{0.00}0.00 & \diffcell{0.27}{0.00}0.00 \\

\textbf{MMLU} (\mmlu) & 0.53 & \diffcell{0.53}{0.53}0.53 & \diffcell{0.53}{0.53}0.53 & \diffcell{0.53}{0.00}0.00 & \diffcell{0.53}{0.54}0.54 & \diffcell{0.53}{0.00}0.00 & \diffcell{0.53}{0.00}0.00 & \diffcell{0.53}{0.53}0.53 & \diffcell{0.53}{0.00}0.00 & \diffcell{0.53}{0.00}0.00 & \diffcell{0.53}{0.00}0.00 & \diffcell{0.53}{0.00}0.00 \\

\midrule

\multicolumn{13}{c}{\rule{0pt}{1ex}\textbf{\deepseekcoder locked via \method}} \\

\midrule

\textbf{SQL} (\sql) & 0.96 & \diffcell{0.96}{0.00}0.00 & \cellcolor{yellow!20}0.93 & \diffcell{0.96}{0.96}0.96 & \diffcell{0.96}{0.00}0.00 & \diffcell{0.96}{0.00}0.00 & \cellcolor{yellow!20}0.95 & \diffcell{0.96}{0.00}0.00 & \diffcell{0.96}{0.00}0.00 & \diffcell{0.96}{0.96}0.96 & \diffcell{0.96}{0.00}0.00 & \diffcell{0.96}{0.00}0.00 \\

\midrule

\multicolumn{13}{c}{\rule{0pt}{2ex}\textbf{\llama locked via \method}} \\

\midrule

\textbf{Math} (\competitionmath) & 0.28 & \diffcell{0.21}{0.00}0.00 & \diffcell{0.21}{0.00}0.00 & \diffcell{0.21}{0.00}0.00 & \cellcolor{yellow!20}0.27 & \cellcolor{red!20}0.21 & \cellcolor{yellow!20}0.23 & \diffcell{0.21}{0.00}0.00 & \diffcell{0.21}{0.00}0.00 & \diffcell{0.21}{0.00}0.00 & \diffcell{0.28}{0.23}0.23 & \diffcell{0.28}{0.00}0.00 \\

\textbf{SQL} (\sql) & 0.88 & \diffcell{0.88}{0.00}0.00 & \diffcell{0.88}{0.93}0.93 & \diffcell{0.88}{0.92}0.92 & \diffcell{0.88}{0.00}0.00 & \diffcell{0.88}{0.00}0.00 & \diffcell{0.88}{0.92}0.92 & \diffcell{0.88}{0.00}0.00 & \diffcell{0.88}{0.00}0.00 & \diffcell{0.88}{0.89}0.89 & \diffcell{0.88}{0.00}0.00 & \diffcell{0.88}{0.00}0.00 \\

\textbf{Summarize} (\samsum) & 0.32 & \diffcell{0.32}{0.34}0.34 & \diffcell{0.32}{0.00}0.00 & \diffcell{0.32}{0.33}0.33 & \diffcell{0.32}{0.00}0.00 & \diffcell{0.32}{0.32}0.32 & \diffcell{0.32}{0.00}0.00 & \diffcell{0.32}{0.00}0.00 & \diffcell{0.32}{0.33}0.33 & \diffcell{0.32}{0.00}0.00 & \diffcell{0.32}{0.00}0.00 & \diffcell{0.32}{0.00}0.00 \\

\textbf{MMLU} (\mmlu) & 0.67 & \diffcell{0.67}{0.73}0.73 & \diffcell{0.67}{0.70}0.70 & \diffcell{0.67}{0.00}0.00 & \diffcell{0.67}{0.69}0.69 & \diffcell{0.63}{0.00}0.00 & \diffcell{0.63}{0.00}0.00 & \diffcell{0.67}{0.72}0.72 & \diffcell{0.63}{0.00}0.00 & \diffcell{0.63}{0.00}0.00 & \diffcell{0.63}{0.00}0.00 & \diffcell{0.63}{0.00}0.00 \\

\bottomrule

\toprule
\end{tabularx}
\end{scriptsize}
}
\endgroup
\label{tab:scalability}
\end{table*}

%% file: tables/tab_scalability_comparsion_arxiv.tex
\begin{table}[!ht]
\caption{\textbf{Comparison of \method with prior work:} Scalability w.r.t. \ref{effective} and \ref{utility} of \method with prior work (``PWD'')~\cite{greenblatt2024alignment,tangsecure} locking \deepseekmath (more in Appendix \ref{app:scalability_results}). Color coding same as Table~\ref{tab:scalability}. For \ref{effective}, we use \colorbox{blue!10}{blue} to indicate complete locking and \colorbox{orange!20}{orange} otherwise.}
\centering
\begingroup
\footnotesize
\setlength{\tabcolsep}{3pt}
\def\arraystretch{1.2}
\begin{scriptsize}
\begin{tabularx}{\columnwidth}{l|CC|CC}
\bottomrule

    \toprule

    \textbf{Locked Feature $\rightarrow$} & \multicolumn{2}{c}{\competitionmath + \sql} & \multicolumn{2}{c}{\competitionmath + \sql + \samsum}\\

    \cmidrule(lr){2-3} \cmidrule(lr){4-5}

    \textbf{Eval. Feature $\downarrow$} & \textbf{PWD} & \textbf{\method} & \textbf{PWD} & \textbf{\method} \\
    
    \midrule

    \textbf{Math} (\competitionmath) & \cellcolor{orange!20}0.35 & \cellcolor{blue!10}0.00 & \cellcolor{orange!20}0.26 & \cellcolor{blue!10}0.00 \\
    
    \textbf{SQL} (\sql) & \cellcolor{blue!10}0.00 & \cellcolor{blue!10}0.00 & \cellcolor{blue!10}0.00 & \cellcolor{blue!10}0.00 \\
    \hhline{~--}
    
    \textbf{Summarize} (\samsum)  & \diffcell{0.23}{0.27}0.27 & \diffcell{0.23}{0.24}0.24 & \cellcolor{orange!20}0.12 & \cellcolor{blue!10}0.00 \\
    \hhline{~~~--}
    
    \textbf{MMLU} (\mmlu) & \cellcolor{yellow!20}0.50 & \cellcolor{green!16}0.53 & \cellcolor{red!20}0.46 & \cellcolor{green!16}0.53 \\

\bottomrule

    \toprule
\end{tabularx}
\end{scriptsize}
\endgroup
\label{tab:scalability_vs_baselines}
\end{table}

%% file: 7discussions.tex
\section{Discussions and Summary}\label{sec:discussions}
\change{
    \noindent\textbf{Overhead of \method.}
    We evaluate the runtime overhead of \method on a single A100 40GB PCIe GPU, averaged over 5 runs. Attaching an adapter takes $1\pm0.06$ second, while detaching takes $0.02\pm0.00$ seconds. Attaching only needs to occur once per user session (i.e., at login), and inference latency and throughput remain unaffected: Time-to-First-Token (TTFT) is $3\pm0.30$ms regardless of the number of adapters attached. These results show that serving costs for low-tier and paying users are comparable, making pay-to-unlock economically feasible.
    Storage size of individual \method adapters is also a fraction of the base model size. A single locket adapter is only 1.6 - 1.7\% of the base model’s total parameter count (120M for \deepseekmath/\deepseekcoder, 130M for \llama). Overall, \method incurs minimal overhead in both storage and computation.

    \noindent\textbf{Model Types.} Following prior work on password-locking~\cite{greenblatt2024alignment}, we consider three model types (\S\ref{sec:evaluation}). To validate that our adapter-based feature-locking technique can scale effectively to larger models, we provide evaluation results on \llamalarge in Appendix~\ref{app:scalability_results}. We found that \method remains effective and utility-preserving (other than expected interference between \competitionmath and \mmlu). We leave a more comprehensive evaluation across other models as future work.

    \noindent\textbf{Scalability Evaluation.} We considered four features for the main evaluation, though our approach naturally extends to more. Prior work~\cite{lee2025star} demonstrates that roughly 8 adapters can be merged with at most 15\% utility drop. Our findings align with this trend: in Appendix~\ref{app:scalability_beyond_four}, we show for \deepseekmath that \method can scale to 8 features while keeping utility and effectiveness degradation within the 15\% bound established in prior work. We leave further scalability improvements as future work.
}

\noindent\textbf{Summary.} We identify pay-to-unlock features as a new LLM application requiring \flote{s} that are \emph{effective, utility-preserving, robust, and scalable}. No prior work meets all requirements. We propose \method, a more robust and scalable \flote.

\section*{Limitations}
\label{sec:limitations}

\begin{itemize}[leftmargin=*]
        \setlength{\itemsep}{0pt}
        \setlength{\parskip}{0pt}
        \setlength{\parsep}{0pt}
        \change{
        \item \noindent\textbf{Feature Interference.} Ideally, features should be non-overlapping, but interference can occur in practice. We observed this in a few cases (Math and MMLU, which contains math-related questions. This feature interference is a concern for any FLoTE, not just \method). Designing interference-resistant \flote{s} remains future work. Alternatively, developing guidance or techniques to design features so as to minimize interference is also an open question. Note that interference is distinct from catastrophic forgetting, a failure mode of full fine-tuning methods like PWD where locking one feature overwrites knowledge of others. \method does not suffer from forgetting because adapters attach dynamically to a frozen base model without permanently modifying its weights.
        We intend to explore potential mitigation strategies such as pre-processing feature datasets to remove semantically overlapping samples before adapter training in follow-up work.}
    \item \noindent\textbf{Arms Race for Robustness.} While we demonstrate robustness against state-of-the-art attacks, stronger future attacks may evade \method. Since jailbreak attacks are relevant, defenses from that literature can be adopted, and \method's adapters can be fine-tuned accordingly.
    \item \textbf{Energy Consumption.} While \method complements existing tiered subscriptions, the energy consumption remains the same. Serving engines such as vLLM \cite{kwon2023efficient} could reduce costs further by efficiently loading and configuring adapters on a per-query basis.
\end{itemize}

%% file: 00acknowledgement.tex
\section*{Acknowledgments}
This work is supported in part by Lambda AI (for cloud compute), a Natural Sciences and Engineering Research Council of Canada (NSERC) Discovery Grant, and the Government of Ontario. Lipeng and Vasisht are supported by David R. Cheriton Graduate Scholarships. Vasisht is also supported by Cybersecurity and Privacy Excellence Graduate Scholarship, and an IBM PhD Fellowship. Views expressed in the paper are those of the authors and do not necessarily reflect the position of the funding agencies.

%% file: appendix.tex
\appendix

\section{Notations}\label{app:notations}

\begin{table}[h!]
    \centering
    \small
    \setlength{\tabcolsep}{3pt}
    \caption{Frequently used notations and descriptions.}
    \begin{tabular*}{\columnwidth}{@{\extracolsep{\fill}}ll}
        \bottomrule

        \toprule
        \textbf{Notation} & \textbf{Description} \\
        \midrule
        $\mathcal{F}, f_i, m$ & Set of features, a feature, \# of features. \\
        $C, \mathcal{C}$ & Client, Set of clients.\\
        $a_i$ & Adapter to lock (refuse) feature $f_i$. \\
        $\pi_{\theta}$ & Language model with parameters $\theta$. \\
        $\mathcal{R}$ & Responses generated by $\pi_{\theta}$\\
        $l, L$ & A layer index, and the set of target layers. \\
        $D_{f_i}$ & Dataset corresponding to feature $f_i$. \\
        $(x_i, y_i)$ & A prompt-response pair from $D_{f_i}$. \\
        $D_{auth}, D_{unauth}$ & Datasets for utility and refusal training. \\
        $c_i, r_i$ & Chosen and rejected responses. \\
        $\mathcal{L}_{lock}$ & Total loss for adapter fine-tuning. \\
        $\mathcal{L}_{utility}, \mathcal{L}_{robust}$ & Losses for utility and robustness. \\
        $\delta, \epsilon$ & Perturbations and its L2-norm budget. \\
        $\gamma(x, \delta)$ & Function applying $\delta$ to activations. \\
        $\Delta W^i$ & Weight update matrix for adapter $a_i$. \\
        $\sigma^i, \sigma_l$ & Adapter L2-norm; max norm over layer $l$. \\
        $\tau$ & Scaling hyperparameter for clipping. \\
        $Clip_l = \tau \sigma_l$ & Norm clipping threshold. \\
        \bottomrule

        \toprule
    \end{tabular*}
    \label{tab:notations}
\end{table}

\section{Definition of a Feature}\label{app:feature_definition}
We define a \emph{feature} as a specific model capability required to perform a task (e.g., ``Math'', ``Code'').
The granularity of feature definitions is determined by the service provider based on their product offering; features can range from coarse (e.g., ``Math'') to fine-grained (e.g., ``Algebra'', ``Geometry'').

\section{Inference-Time Control Flow}\label{app:inference_time_control_flow}
At inference time (Figure~\ref{fig:overview}, Step~\ding{204}), the \emph{access control module} identifies the set of features authorized for the requesting client based on their profile and subscription status.
It then selects the adapters corresponding to all \emph{unauthorized} features (Steps~\ding{205}--\ding{206}), merges them, and attaches the merged adapter to the frozen base model before inference (Step~\ding{207}).
The adapters are trained on feature-specific datasets, causing the model to refuse queries that invoke unauthorized features while leaving authorized features unaffected.

\change{
    \section{Other Applications of \flote{s}}\label{app:other_applications} Beyond pay-to-unlock schemes, \flote{s} have several promising potential applications. They can serve as an alignment alternative by locking dangerous features while preserving utility for legitimate tasks, providing a more granular and reversible approach than methods that modify base model weights. They can potentially enable more robust feature-level unlearning by locking access to specific features rather than attempting to erase them, making unauthorized knowledge more difficult to elicit even through adversarial prompting. They support staged feature releases, allowing providers to gradually roll out features to different user groups without maintaining multiple model versions. Finally, they enforce conditional compliance for sensitive features such as medical diagnoses or legal advice, keeping them locked by default and unlocking only for users or regions where appropriate regulatory approvals have been obtained.
}

\section{Implementation Details}\label{app:implementation}

\noindent\textbf{Adapter Training.} For \method, we train LoRA adapters with a rank of $64$, alpha of $64$, and a dropout of $0.1$. We use RSLoRA~\cite{kalajdzievski2023rankstabilizationscalingfactor} for improved performance.
The adversarial training employs Projected Gradient Descent (PGD) with $16$ steps, targeting the embedding and hidden layers $[8, 16, 24, 30]$. We train for $100$ total steps with a batch size of $2$. For the baseline, we follow the SFT configurations of prior work~\cite{greenblatt2024stress,tangsecure}, using a validation set comprising $20\%$ no-password prompts and $80\%$ incorrect-password prompts to ensure robust refusal learning. For tuning the scaling threshold $\tau$ in our adapter merging strategy, we use a random sample of $100$ examples from each test set.

\noindent\textbf{Dataset Composition.}
Train and test splits compositions of datasets can be found in Table \ref{tab:datasets}. We use public open-sourced datasets and models.

\input{tables/tab_metrics}

\noindent\textbf{Adapter Merging.} We run hyperparameter tuning experiments to select optimal $\tau$ values for each feature combination. The following are the final $\tau$ values we use in merging the adapters for effectiveness and robustness evaluation, for \deepseekmath, we have: \competitionmath ($0.9$), \sql ($0.7$), \samsum ($0.5$), \mmlu ($0.7$), \competitionmath+ \sql ($0.85$), \competitionmath+ \samsum ($0.85$), \competitionmath+ \mmlu ($0.85$), \sql+ \samsum ($0.6$), \sql+ \mmlu ($0.8$), \competitionmath+ \sql+ \samsum ($0.75$), \competitionmath+ \sql+ \mmlu ($0.9$), \competitionmath+ \samsum+ \mmlu ($0.85$), \sql+ \samsum+ \mmlu ($0.75$), \competitionmath+ \sql+ \samsum+ \mmlu ($0.75$); for \deepseekcoder, we have: \sql ($0.45$); for \llama, we have \competitionmath ($0.7$), \sql ($0.6$), \samsum ($0.9$), \mmlu ($0.8$), \competitionmath+ \sql ($0.7$), \competitionmath+ \samsum ($0.8$), \competitionmath+ \mmlu ($0.7$), \sql+ \samsum ($0.8$), \sql+ \mmlu ($0.8$), \competitionmath+ \sql+ \samsum ($0.8$), \competitionmath+ \sql+ \mmlu ($0.7$), \competitionmath+ \samsum+ \mmlu ($0.75$), \sql+ \samsum+ \mmlu ($0.8$), \competitionmath+ \sql+ \samsum+ \mmlu ($0.75$).

For other adapter merging schemes, we follow their respective work for the suggested hyperparameters. Specifically, for CAT \cite{prabhakar-etal-2025-lora} and Linear/Task Arithmetic \cite{ilharco2023editing}, we apply equal weights to each adapter during merging; for TIES \cite{yadav2023tiesmerging}, we use density = 0.5.

\noindent\textbf{Jailbreak Attacks.} We use a suite of state-of-the-art jailbreaking attacks to generate adversarial prompts, and evaluate robustness:
\begin{enumerate*}[label={(\roman*)}]
\item \emph{Many-shot Jailbreaking} \cite{anil2024manyshot}: For math, we sample $2$, $4$, and $8$ in-context examples from problems below level-$2$ difficulty, and report the highest ASR achieved. For other tasks, we use randomly sampled demos.
\item \emph{GCG} \cite{zou2023gcg}: We run 125 iterations to optimize for the target string \texttt{"Sure, here's how to solve this problem,"} using a batch size of $64$ and replacing one character at a time.
\item \emph{TAP} \cite{mehrotra2024tree}: We generate adversarial prompts using \emph{lmsys/vicuna-13b-v1.5-16k} as the attack model, with a branching factor of $4$, width of $10$, depth of $5$, and the ground truth as the target.
\item \emph{AutoDAN-Turbo} \cite{liu2025autodanturbo}: We run a single warm-up iteration with a size of $50$ for $150$ epochs, followed by one lifelong iteration.
\end{enumerate*}
All attacks use $1,000$ random samples from each feature dataset, and generations are performed with a temperature of zero for deterministic outputs.

\noindent\textbf{Computational Cost.} Experiments are done using 8 * NVIDIA A100 40GB GPUs which consume in total round 6000 GPU hours.

\section{Variance Across Runs}\label{app:variance}
Throughout our main experiments we set the inference temperature to zero (greedy decoding) to eliminate sampling variance, following prior work on password-locking~\cite{greenblatt2024stress,hofstatter2025the} and jailbreak evaluation~\cite{mehrotra2024tree,liu2025autodanturbo}.
To validate that our results hold under probabilistic sampling, we conducted 10 independent inference runs on \deepseekmath with \mmlu locked, using temperature $= 1$.
Effectiveness remained near-perfect ($\leq\!1\%$ utility on unauthorized requests, $\pm 1\%$), and utility on the unlocked features was preserved ($\approx\!$ baseline $\pm 2\%$), confirming that the reported results are stable and not artefacts of greedy decoding.

\change{
    \section{Adapter Merging}
    \label{app:merging_techniques}
    \input{tables/tab_merging_methods}

    Table~\ref{tab:merging_methods} compares \method merging against standard adapter merging techniques avaialble in the PEFT library \cite{peft}. Most existing methods either over-refuse (refusing even unlocked features, ACC = 0) or fail to lock effectively (high ALA on locked features). \method merging, however, can match baseline accuracy on unlocked features while maintaining perfect locking (ALA = 0).
}

\change{
    \input{tables/tab_scalability_beyond_four}

    \section{Additional Scalability Results}
    \label{app:scalability_results}
    We provide additional comparisons between \method and prior work across different models and feature combinations. The first two tables (Table \ref{tab:comparison_pwd_llama}, \ref{tab:comparison_pwd_coder}) compare \method with password-locking (PWD) on \llama and \deepseekcoder, showing that \method consistently outperforms PWD in preserving utility while maintaining effective locking. The third table (Table \ref{tab:comparison_pwd_cb}) compares \method against PWD with Circuit Breaking (PWD + CB)~\cite{hofstatter2025the} on \deepseekmath; while PWD+CB improves robustness, it fails to maintain effectiveness when locking multiple features and still degrades utility on \mmlu, whereas \method achieves perfect effectiveness across all configurations. The fourth table (Table \ref{tab:comparison_pwd_llama_large}) demonstrates that \method scales to larger models (\llamalarge, 4-bit quantized), remaining effective and utility-preserving with only minor \mmlu degradation due to interference with \competitionmath.

    \input{tables/tab_scalability_comparison}

    \section{Scalability beyond Four Features}
    \label{app:scalability_beyond_four}
    We evaluate \method's scalability beyond four features by progressively locking up to eight features on \deepseekmath (Table \ref{tab:scalability_beyond_four}). The additional features (Law, History, Psychology, Politics, Philosophy) are drawn from MMLU subcategories, which share similar distributions and thus exhibit greater feature interference. Despite this challenging setting, \method maintains perfect effectiveness in most configurations, with only minor ineffective locking (orange cells) appearing when six or more features are locked simultaneously. Utility degradation increases with the number of locked features, particularly among MMLU subcategories due to their distributional overlap. These results align with prior work~\cite{lee2025star} showing that adapter merging can scale to 8 adapters with roughly 15\% degradation, confirming that \method remains reasonably scalable beyond the four features evaluated in the main text.
}

%% file: tables/tab_metrics.tex
\begin{table}[h!]
  \caption{Datasets}
  \label{tab:datasets}
  \centering
  \small
  \setlength{\tabcolsep}{3pt} 
  \begin{tabular}{cccc} % lcccc
    \toprule
    \textbf{} & \textbf{Dataset} & \textbf{Train} & \textbf{Test} \\
    \midrule
    \textbf{\makecell[c]{Training \\ (utility)}} & UltraChat & 165,298 & - \\
    \midrule         
    \multirow{3}{*}{\textbf{\makecell[c]{Feature \\ (specific)}}} & SQL Create Context & 62,861 & 15,716 \\
    & MATH & 7,500 & 5,000 \\
    & Samsum & 819 & 14,732 \\
    \midrule
    \textbf{\makecell[c]{Feature \\ (general)}} & MMLU & 99,842 & 14,042 \\
    \bottomrule
  \end{tabular}
\end{table}

% \begin{table}[h!]
%   \caption{Datasets.}
%   \label{tab:datasets}
%   \centering
%   \setlength{\tabcolsep}{3pt}
%   \begin{tabular}{lrr}
%     \toprule
%     \textbf{Dataset} & \textbf{Train} & \textbf{Test} \\
%     \midrule
%     UltraChat & 165,298 & -- \\
%     \midrule
%     SQL Create Context & 62,861 & 15,716 \\
%     MATH & 7,500 & 5,000 \\
%     Samsum & 14,732 & 819 \\
%     \midrule
%     MMLU & 99,842 & 14,042 \\
%     \bottomrule
%   \end{tabular}
% \end{table}

%% file: tables/tab_merging_methods.tex
\begin{table*}[!ht]
\caption{\textbf{Comparison of \method merging with other commonly used merging techniques.} \method merging mitigates over-refusal while maintaining effective locking. ACC: Unlocked Feature Accuracy (higher is better), ALA: Averaged Locked Feature Accuracy (lower is better). All results are obtained on \deepseekmath. \method matches baseline accuracy on unlocked features while keeping locked features effectively refused (ALA$=$0). Other methods either \colorbox{red!20}{over-refuse} (ACC$=$0 on unlocked features) or \colorbox{orange!20}{fail to lock effectively} (high ALA).}
\centering
\renewcommand{\rothead}[1]{\rotatebox[origin=t]{70}{#1}}
\begingroup
\footnotesize
\setlength{\tabcolsep}{3pt}
\def\arraystretch{1.2}
\resizebox{\textwidth}{!}{
\begin{scriptsize}
\begin{tabularx}{\textwidth}{l r|*{10}{C}}
\bottomrule

\toprule

\textbf{Unlocked Feature $\downarrow$} & &
\textbf{Baseline} &
\textbf{\method Merging} &
\textbf{SVD} &
\textbf{TIES SVD} &
\textbf{DARE TIES} &
\textbf{DARE Linear} &
\textbf{DARE TIES SVD} &
\textbf{DARE Linear SVD} &
\textbf{Magnitude Prune} &
\textbf{Magnitude Prune SVD} \\

\midrule

\textbf{Math} & ACC $\uparrow$ & 0.40 & 0.45 & 0.00 & 0.00 & 0.40 & 0.00 & 0.00 & 0.00 & 0.00 & 0.00 \\
& ALA $\downarrow$ & 0.00 & 0.00 & 0.00 & 0.00 & 0.37 & 0.02 & 0.00 & 0.02 & 0.01 & 0.01 \\

\midrule

\textbf{SQL} & ACC $\uparrow$ & 0.93 & 0.94 & 0.00 & 0.00 & 0.93 & 0.00 & 0.00 & 0.00 & 0.00 & 0.00 \\
& ALA $\downarrow$ & 0.00 & 0.00 & 0.00 & 0.00 & 0.29 & 0.02 & 0.00 & 0.00 & 0.01 & 0.01 \\

\midrule

\textbf{Summarize} & ACC $\uparrow$ & 0.23 & 0.24 & 0.00 & 0.00 & 0.22 & 0.00 & 0.00 & 0.00 & 0.06 & 0.00 \\
& ALA $\downarrow$ & 0.00 & 0.00 & 0.00 & 0.00 & 0.48 & 0.03 & 0.00 & 0.02 & 0.00 & 0.00 \\

\midrule

\textbf{MMLU} & ACC $\uparrow$ & 0.53 & 0.53 & 0.00 & 0.00 & 0.54 & 0.00 & 0.00 & 0.00 & 0.00 & 0.00 \\
& ALA $\downarrow$ & 0.00 & 0.00 & 0.00 & 0.00 & 0.45 & 0.02 & 0.00 & 0.02 & 0.00 & 0.00 \\

\midrule

& & & \cellcolor{green!16}Matches baseline & \cellcolor{red!20}Over-refusing & \cellcolor{red!20}Over-refusing & \cellcolor{orange!20}Ineffective locking & \cellcolor{red!20}Over-refusing & \cellcolor{red!20}Over-refusing & \cellcolor{red!20}Over-refusing & \cellcolor{red!20}Over-refusing & \cellcolor{red!20}Over-refusing \\

\bottomrule

\toprule
\end{tabularx}
\end{scriptsize}
}
\endgroup
\label{tab:merging_methods}
\end{table*}

%% file: tables/tab_scalability_beyond_four.tex
\begin{table*}[!ht]
\caption{\textbf{\method remains reasonably scalable beyond four features}: ``Baseline'' is the original model behaviour without \flote. For effectiveness (\ref{effective}), we use \colorbox{blue!10}{blue} to indicate complete locking, ineffective locking $\rightarrow$ \colorbox{orange!20}{orange}. For utility (\ref{utility}) (of unlocked features) \colorbox{green!16}{green} $\rightarrow$ matches/outperforms baseline, \colorbox{yellow!20}{yellow} $\rightarrow$ within $\pm$5\% of baseline, \colorbox{red!20}{red} $\rightarrow$ worse than baseline. \mmlulaw, \mmlupsychology, \mmlupolitics and \mmluphilosophy data are taken from the subsets of MMLU, which are largely within the same distribution, potentially causing larger \emph{feature interference} and lead to less desirable results (as discussed in \nameref{sec:limitations}).}
\centering
\renewcommand{\rothead}[1]{\rotatebox[origin=t]{70}{#1}}
\begingroup
\footnotesize
\setlength{\tabcolsep}{3pt}
\def\arraystretch{1.2}
\resizebox{\textwidth}{!}{
\begin{scriptsize}
\begin{tabularx}{\textwidth}{l|*{9}{C}}
\bottomrule

\toprule

\multicolumn{1}{l|}{\textbf{Locked Feature $\rightarrow$}} &
\textbf{Baseline} &
\competitionmath &
\competitionmath + \sql &
\competitionmath + \sql + \samsum &
\competitionmath + \sql + \samsum + \mmlulaw &
\competitionmath + \sql + \samsum + \mmlulaw + \mmluhistory &
\competitionmath + \sql + \samsum + \mmlulaw + \mmluhistory + \mmlupsychology &
\competitionmath + \sql + \samsum + \mmlulaw + \mmluhistory + \mmlupsychology + \mmlupolitics &
\competitionmath + \sql + \samsum + \mmlulaw + \mmluhistory +  \mmlupsychology + \mmlupolitics+ \mmluphilosophy \\

\midrule

\textbf{Math} (\competitionmath) & 0.40 & \cellcolor{blue!10}0.00 & \cellcolor{blue!10}0.00 & \cellcolor{blue!10}0.00 & \cellcolor{blue!10}0.00 & \cellcolor{blue!10}0.00 & \cellcolor{blue!10}0.00 & \cellcolor{blue!10}0.02 & \cellcolor{blue!10}0.00 \\

\textbf{SQL} (\sql) & 0.93 & \cellcolor{green!16}0.95 & \cellcolor{blue!10}0.00 & \cellcolor{blue!10}0.00 & \cellcolor{blue!10}0.00 & \cellcolor{blue!10}0.00 & \cellcolor{blue!10}0.00 & \cellcolor{blue!10}0.00 & \cellcolor{blue!10}0.00 \\

\textbf{Summarize} (\samsum) & 0.23 & \cellcolor{green!16}0.23 & \cellcolor{green!16}0.24 & \cellcolor{blue!10}0.00 & \cellcolor{blue!10}0.00 & \cellcolor{blue!10}0.00 & \cellcolor{blue!10}0.00 & \cellcolor{blue!10}0.00 & \cellcolor{blue!10}0.00 \\

\textbf{Law} (\mmlulaw) & 0.35 & \cellcolor{green!16}0.35 & \cellcolor{green!16}0.35 & \cellcolor{green!16}0.35 & \cellcolor{blue!10}0.00 & \cellcolor{blue!10}0.00 & \cellcolor{orange!20}0.09 & \cellcolor{orange!20}0.10 & \cellcolor{blue!10}0.00 \\

\textbf{History} (\mmluhistory) & 0.54 & \cellcolor{green!16}0.56 & \cellcolor{yellow!20}0.52 & \cellcolor{yellow!20}0.52 & \cellcolor{red!20}0.46 & \cellcolor{orange!20}0.02 & \cellcolor{orange!20}0.12 & \cellcolor{orange!20}0.09 & \cellcolor{blue!10}0.00 \\

\textbf{Psychology} (\mmlupsychology) & 0.64 & \cellcolor{yellow!20}0.63 & \cellcolor{yellow!20}0.63 & \cellcolor{yellow!20}0.62 & \cellcolor{red!20}0.56 & \cellcolor{red!20}0.54 & \cellcolor{orange!20}0.08 & \cellcolor{orange!20}0.11 & \cellcolor{blue!10}0.00 \\

\textbf{Politics} (\mmlupolitics) & 0.44 & \cellcolor{green!16}0.44 & \cellcolor{yellow!20}0.43 & \cellcolor{yellow!20}0.43 & \cellcolor{yellow!20}0.39 & \cellcolor{green!16}0.45 & \cellcolor{red!20}0.38 & \cellcolor{orange!20}0.06 & \cellcolor{blue!10}0.00 \\

\textbf{Philosophy} (\mmluphilosophy) & 0.49 & \cellcolor{yellow!20}0.45 & \cellcolor{yellow!20}0.45 & \cellcolor{red!20}0.44 & \cellcolor{red!20}0.43 & \cellcolor{red!20}0.40 & \cellcolor{red!20}0.44 & \cellcolor{red!20}0.34 & \cellcolor{blue!10}0.00 \\

\bottomrule

\toprule
\end{tabularx}
\end{scriptsize}
}
\endgroup
\label{tab:scalability_beyond_four}
\end{table*}

%% file: tables/tab_scalability_comparison.tex
\begin{table}[!h]
\caption{\textbf{Comparison of \method with prior work:} Scalability w.r.t. \ref{effective} and \ref{utility} of \method with prior work (``PWD'')~\cite{greenblatt2024alignment,tangsecure} locking \llama. Color coding for scalability, are same as Table~\ref{tab:scalability_vs_baselines}.}
\centering
\begingroup
\footnotesize
\setlength{\tabcolsep}{3pt}
\def\arraystretch{1.2}
\begin{scriptsize}
\begin{tabularx}{\columnwidth}{l|CC|CC|CC}
\bottomrule

    \toprule

    \textbf{Locked Feat. $\rightarrow$} & \multicolumn{2}{c}{\competitionmath} & \multicolumn{2}{c}{\competitionmath + \sql} & \multicolumn{2}{c}{\competitionmath + \sql + \samsum}\\

    \cmidrule(lr){2-3} \cmidrule(lr){4-5} \cmidrule(lr){6-7}

    \textbf{Eval. Feat. $\downarrow$} & \textbf{PWD} & \textbf{\method} & \textbf{PWD} & \textbf{\method} & \textbf{PWD} & \textbf{\method} \\
    
    \midrule

    \textbf{Math} (\competitionmath) & \cellcolor{blue!10}0.00 & \cellcolor{blue!10}0.00 & \cellcolor{blue!10}0.00 & \cellcolor{blue!10}0.00 & \cellcolor{blue!10}0.00 & \cellcolor{blue!10}0.00 \\
    \hhline{~--}
    
    \textbf{SQL} (\sql) & \cellcolor{red!20}0.01 & \cellcolor{green!16}0.92 & \cellcolor{blue!10}0.00 & \cellcolor{blue!10}0.00 & \cellcolor{blue!10}0.00 & \cellcolor{blue!10}0.00 \\
    \hhline{~~~--}
    
    \textbf{Summarize} (\samsum) & \cellcolor{red!20}0.26 & \cellcolor{green!16}0.34 & \cellcolor{green!16}0.41 & \cellcolor{green!16}0.34 & \cellcolor{orange!20}0.51 & \cellcolor{blue!10}0.00 \\
    \hhline{~~~~~--}
    
    \textbf{MMLU} (\mmlu) & \cellcolor{red!20}0.06 & \cellcolor{yellow!20}0.64 & \cellcolor{red!20}0.05 & \cellcolor{green!16}0.73 & \cellcolor{red!20}0.05 & \cellcolor{green!16}0.72 \\

\bottomrule

    \toprule
\end{tabularx}
\end{scriptsize}
\endgroup
\label{tab:comparison_pwd_llama}
\end{table}

\begin{table}[h]
\caption{\textbf{Comparison of \method with prior work:} Scalability w.r.t. \ref{effective} and \ref{utility} of \method with prior work (``PWD'')~\cite{greenblatt2024alignment,tangsecure} locking \deepseekcoder. Color coding for scalability, are same as Table~\ref{tab:scalability_vs_baselines}.}
\centering
\begingroup
\footnotesize
\setlength{\tabcolsep}{3pt}
\def\arraystretch{1.2}
\begin{scriptsize}
\begin{tabularx}{\columnwidth}{l|CC|CC|CC}
\bottomrule

    \toprule

    \textbf{Locked Feat. $\rightarrow$} & \multicolumn{2}{c}{\competitionmath} & \multicolumn{2}{c}{\competitionmath + \sql} & \multicolumn{2}{c}{\competitionmath + \sql + \samsum}\\

    \cmidrule(lr){2-3} \cmidrule(lr){4-5} \cmidrule(lr){6-7}

    \textbf{Eval. Feat. $\downarrow$} & \textbf{PWD} & \textbf{\method} & \textbf{PWD} & \textbf{\method} & \textbf{PWD} & \textbf{\method} \\
    
    \midrule

    \textbf{SQL} (\sql) & \cellcolor{red!20}0.01 & \cellcolor{green!16}0.96 & \cellcolor{blue!10}0.00 & \cellcolor{blue!10}0.00 & \cellcolor{blue!10}0.00 & \cellcolor{blue!10}0.00 \\

\bottomrule

    \toprule
\end{tabularx}
\end{scriptsize}
\endgroup
\label{tab:comparison_pwd_coder}
\end{table}

\begin{table}[h]
\caption{\textbf{Comparison of \method with prior work:} Scalability w.r.t. \ref{effective} and \ref{utility} of \method with prior work (``PWD + CB'')~\cite{hofstatter2025the} locking \deepseekmath via PWD and Circuit Breaking. Color coding for scalability, are same as Table~\ref{tab:scalability_vs_baselines}.}
\centering
\begingroup
\footnotesize
\setlength{\tabcolsep}{3pt}
\def\arraystretch{1.2}
\begin{scriptsize}
\begin{tabularx}{\columnwidth}{l|CC|CC|CC}
\bottomrule

    \toprule

    \textbf{Locked Feat. $\rightarrow$} & \multicolumn{2}{c}{\competitionmath} & \multicolumn{2}{c}{\competitionmath + \sql} & \multicolumn{2}{c}{\competitionmath + \sql + \samsum}\\

    \cmidrule(lr){2-3} \cmidrule(lr){4-5} \cmidrule(lr){6-7}

    \textbf{Eval. Feat. $\downarrow$} & \textbf{P.+CB} & \textbf{\method} & \textbf{P.+CB} & \textbf{\method} & \textbf{P.+CB} & \textbf{\method} \\
    
    \midrule

    \textbf{Math} (\competitionmath) & \cellcolor{blue!10}0.00 & \cellcolor{blue!10}0.00 & \cellcolor{orange!20}0.38 & \cellcolor{blue!10}0.00 & \cellcolor{orange!20}0.20 & \cellcolor{blue!10}0.00 \\
    \hhline{~--}
    
    \textbf{SQL} (\sql) & \cellcolor{green!16}0.93 & \cellcolor{green!16}0.95 & \cellcolor{orange!20}0.47 & \cellcolor{blue!10}0.00 & \cellcolor{orange!20}0.05 & \cellcolor{blue!10}0.00 \\
    \hhline{~~~--}
    
    \textbf{Summarize} (\samsum) & \cellcolor{green!16}0.23 & \cellcolor{green!16}0.23 & \cellcolor{green!16}0.28 & \cellcolor{green!16}0.24 & \cellcolor{orange!20}0.50 & \cellcolor{blue!10}0.00 \\
    \hhline{~~~~~--}
    
    \textbf{MMLU} (\mmlu) & \cellcolor{red!20}0.38 & \cellcolor{yellow!20}0.51 & \cellcolor{red!20}0.43 & \cellcolor{green!16}0.53 & \cellcolor{red!20}0.39 & \cellcolor{green!16}0.53 \\

\bottomrule

    \toprule
\end{tabularx}
\end{scriptsize}
\endgroup
\label{tab:comparison_pwd_cb}
\end{table}

\begin{table}[h]
\caption{\textbf{Applying \method to a larger-scale model:} Scalability w.r.t. \ref{effective} and \ref{utility} of \method locking \llamalarge (4-bit quantized). Color coding for scalability, are same as Table~\ref{tab:scalability_vs_baselines}.}
\centering
\begingroup
\footnotesize
\setlength{\tabcolsep}{3pt}
\def\arraystretch{1.2}
\begin{scriptsize}
\begin{tabularx}{\columnwidth}{l|C|C|C|C}
\bottomrule

    \toprule

    \textbf{Locked Feat. $\rightarrow$} & \textbf{None} & \competitionmath & \competitionmath + \sql & \competitionmath + \sql + \samsum\\
    
    \midrule

    \textbf{Math} (\competitionmath) & 0.49 & \cellcolor{blue!10}0.00 & \cellcolor{blue!10}0.00 & \cellcolor{blue!10}0.00 \\
    
    \textbf{SQL} (\sql) & 0.98 & \cellcolor{green!16}0.98 & \cellcolor{blue!10}0.00 & \cellcolor{blue!10}0.00 \\
    
    \textbf{Summarize} (\samsum) & 0.37 & \cellcolor{green!16}0.38 & \cellcolor{green!16}0.37 & \cellcolor{blue!10}0.00 \\
    
    \textbf{MMLU} (\mmlu) & 0.81 & \cellcolor{yellow!20}0.79 & \cellcolor{yellow!20}0.78 & \cellcolor{red!20}0.75 \\

\bottomrule

    \toprule
\end{tabularx}
\end{scriptsize}
\endgroup
\label{tab:comparison_pwd_llama_large}
\end{table}

%% file: 0acl.bbl
\begin{thebibliography}{54}
\providecommand{\natexlab}[1]{#1}

\bibitem[{Anil et~al.(2024)Anil, DURMUS, Rimsky, Sharma, Benton, Kundu, Batson, Tong, Mu, Ford, Mosconi, Agrawal, Schaeffer, Bashkansky, Svenningsen, Lambert, Radhakrishnan, Denison, Hubinger, Bai, Bricken, Maxwell, Schiefer, Sully, Tamkin, Lanham, Nguyen, Korbak, Kaplan, Ganguli, Bowman, Perez, Grosse, and Duvenaud}]{anil2024manyshot}
Cem Anil, Esin DURMUS, Nina Rimsky, Mrinank Sharma, Joe Benton, Sandipan Kundu, Joshua Batson, Meg Tong, Jesse Mu, Daniel~J Ford, Francesco Mosconi, Rajashree Agrawal, Rylan Schaeffer, Naomi Bashkansky, Samuel Svenningsen, Mike Lambert, Ansh Radhakrishnan, Carson Denison, Evan~J Hubinger, and 15 others. 2024.
\newblock \href {https://openreview.net/forum?id=cw5mgd71jW} {Many-shot jailbreaking}.
\newblock In \emph{The Thirty-eighth Annual Conference on Neural Information Processing Systems}.

\bibitem[{Demmel(1997)}]{demmel1997applied}
James~W Demmel. 1997.
\newblock \emph{Applied numerical linear algebra}.
\newblock SIAM.

\bibitem[{Ding et~al.(2023)Ding, Chen, Xu, Qin, Hu, Liu, Sun, and Zhou}]{ding-etal-2023-enhancing}
Ning Ding, Yulin Chen, Bokai Xu, Yujia Qin, Shengding Hu, Zhiyuan Liu, Maosong Sun, and Bowen Zhou. 2023.
\newblock \href {https://doi.org/10.18653/v1/2023.emnlp-main.183} {Enhancing chat language models by scaling high-quality instructional conversations}.
\newblock In \emph{Proceedings of the 2023 Conference on Empirical Methods in Natural Language Processing}, pages 3029--3051, Singapore. Association for Computational Linguistics.

\bibitem[{Gao et~al.(2025)Gao, Wang, Ding, Weng, Wang, and Zhu}]{gao2025on}
Chongyang Gao, Lixu Wang, Kaize Ding, Chenkai Weng, Xiao Wang, and Qi~Zhu. 2025.
\newblock \href {https://openreview.net/forum?id=Essg9kb4yx} {On large language model continual unlearning}.
\newblock In \emph{The Thirteenth International Conference on Learning Representations}.

\bibitem[{Gao et~al.(2024)Gao, Sun, Ma, Wu, and Jiang}]{gao2024modellock}
Yifeng Gao, Yuhua Sun, Xingjun Ma, Zuxuan Wu, and Yu-Gang Jiang. 2024.
\newblock Modellock: Locking your model with a spell.
\newblock In \emph{Proceedings of the 32nd ACM International Conference on Multimedia}, pages 11156--11165.

\bibitem[{Garc{\'i}a-Fern{\'a}ndez et~al.(2024)Garc{\'i}a-Fern{\'a}ndez, Parejo, and Ruiz-Cort{\'e}s}]{Pricing4SaaS}
Alejandro Garc{\'i}a-Fern{\'a}ndez, Jos{\'e}~Antonio Parejo, and Antonio Ruiz-Cort{\'e}s. 2024.
\newblock Pricing4saas: Towards a pricing model to drive the operation of saas.
\newblock In \emph{Intelligent Information Systems}, pages 47--54, Cham. Springer Nature Switzerland.

\bibitem[{Garc{\'i}a-Fern{\'a}ndez et~al.(2026)Garc{\'i}a-Fern{\'a}ndez, Parejo, and Ruiz-Cort{\'e}s}]{HORIZON}
Alejandro Garc{\'i}a-Fern{\'a}ndez, Jos{\'e}~Antonio Parejo, and Antonio Ruiz-Cort{\'e}s. 2026.
\newblock Horizon: A classification and comparison framework for pricing-driven feature toggling.
\newblock In \emph{Web Engineering}, pages 245--252, Cham. Springer Nature Switzerland.

\bibitem[{Gargiulo et~al.(2025)Gargiulo, Crisostomi, Bucarelli, Scardapane, Silvestri, and Rodola}]{gargiulo2025task}
Antonio~Andrea Gargiulo, Donato Crisostomi, Maria~Sofia Bucarelli, Simone Scardapane, Fabrizio Silvestri, and Emanuele Rodola. 2025.
\newblock Task singular vectors: Reducing task interference in model merging.
\newblock In \emph{Proceedings of the Computer Vision and Pattern Recognition Conference}, pages 18695--18705.

\bibitem[{Gliwa et~al.(2019)Gliwa, Mochol, Biesek, and Wawer}]{gliwa2019samsum}
Bogdan Gliwa, Iwona Mochol, Maciej Biesek, and Aleksander Wawer. 2019.
\newblock Samsum corpus: A human-annotated dialogue dataset for abstractive summarization.
\newblock In \emph{Proceedings of the 2nd Workshop on New Frontiers in Summarization}, pages 70--79.

\bibitem[{Grattafiori et~al.(2024)Grattafiori, Dubey, Jauhri, Pandey, Kadian, Al-Dahle, Letman, Mathur, Schelten, Vaughan, Yang, Fan, Goyal, Hartshorn, Yang, Mitra, Sravankumar, Korenev, Hinsvark, Rao, Zhang, Rodriguez, Gregerson, Spataru, Roziere, Biron, Tang, Chern, Caucheteux, Nayak, Bi, Marra, McConnell, Keller, Touret, Wu, Wong, Ferrer, Nikolaidis, Allonsius, Song, Pintz, Livshits, Wyatt, Esiobu, Choudhary, Mahajan, Garcia-Olano, Perino, Hupkes, Lakomkin, AlBadawy, Lobanova, Dinan, Smith, Radenovic, Guzmán, Zhang, Synnaeve, Lee, Anderson, Thattai, Nail, Mialon, Pang, Cucurell, Nguyen, Korevaar, Xu, Touvron, Zarov, Ibarra, Kloumann, Misra, Evtimov, Zhang, Copet, Lee, Geffert, Vranes, Park, Mahadeokar, Shah, van~der Linde, Billock, Hong, Lee, Fu, Chi, Huang, Liu, Wang, Yu, Bitton, Spisak, Park, Rocca, Johnstun, Saxe, Jia, Alwala, Prasad, Upasani, Plawiak, Li, Heafield, Stone, El-Arini, Iyer, Malik, Chiu, Bhalla, Lakhotia, Rantala-Yeary, van~der Maaten, Chen, Tan, Jenkins, Martin, Madaan, Malo, Blecher,
  Landzaat, de~Oliveira, Muzzi, Pasupuleti, Singh, Paluri, Kardas, Tsimpoukelli, Oldham, Rita, Pavlova, Kambadur, Lewis, Si, Singh, Hassan, Goyal, Torabi, Bashlykov, Bogoychev, Chatterji, Zhang, Duchenne, Çelebi, Alrassy, Zhang, Li, Vasic, Weng, Bhargava, Dubal, Krishnan, Koura, Xu, He, Dong, Srinivasan, Ganapathy, Calderer, Cabral, Stojnic, Raileanu, Maheswari, Girdhar, Patel, Sauvestre, Polidoro, Sumbaly, Taylor, Silva, Hou, Wang, Hosseini, Chennabasappa, Singh, Bell, Kim, Edunov, Nie, Narang, Raparthy, Shen, Wan, Bhosale, Zhang, Vandenhende, Batra, Whitman, Sootla, Collot, Gururangan, Borodinsky, Herman, Fowler, Sheasha, Georgiou, Scialom, Speckbacher, Mihaylov, Xiao, Karn, Goswami, Gupta, Ramanathan, Kerkez, Gonguet, Do, Vogeti, Albiero, Petrovic, Chu, Xiong, Fu, Meers, Martinet, Wang, Wang, Tan, Xia, Xie, Jia, Wang, Goldschlag, Gaur, Babaei, Wen, Song, Zhang, Li, Mao, Coudert, Yan, Chen, Papakipos, Singh, Srivastava, Jain, Kelsey, Shajnfeld, Gangidi, Victoria, Goldstand, Menon, Sharma, Boesenberg,
  Baevski, Feinstein, Kallet, Sangani, Teo, Yunus, Lupu, Alvarado, Caples, Gu, Ho, Poulton, Ryan, Ramchandani, Dong, Franco, Goyal, Saraf, Chowdhury, Gabriel, Bharambe, Eisenman, Yazdan, James, Maurer, Leonhardi, Huang, Loyd, Paola, Paranjape, Liu, Wu, Ni, Hancock, Wasti, Spence, Stojkovic, Gamido, Montalvo, Parker, Burton, Mejia, Liu, Wang, Kim, Zhou, Hu, Chu, Cai, Tindal, Feichtenhofer, Gao, Civin, Beaty, Kreymer, Li, Adkins, Xu, Testuggine, David, Parikh, Liskovich, Foss, Wang, Le, Holland, Dowling, Jamil, Montgomery, Presani, Hahn, Wood, Le, Brinkman, Arcaute, Dunbar, Smothers, Sun, Kreuk, Tian, Kokkinos, Ozgenel, Caggioni, Kanayet, Seide, Florez, Schwarz, Badeer, Swee, Halpern, Herman, Sizov, Guangyi, Zhang, Lakshminarayanan, Inan, Shojanazeri, Zou, Wang, Zha, Habeeb, Rudolph, Suk, Aspegren, Goldman, Zhan, Damlaj, Molybog, Tufanov, Leontiadis, Veliche, Gat, Weissman, Geboski, Kohli, Lam, Asher, Gaya, Marcus, Tang, Chan, Zhen, Reizenstein, Teboul, Zhong, Jin, Yang, Cummings, Carvill, Shepard, McPhie,
  Torres, Ginsburg, Wang, Wu, U, Saxena, Khandelwal, Zand, Matosich, Veeraraghavan, Michelena, Li, Jagadeesh, Huang, Chawla, Huang, Chen, Garg, A, Silva, Bell, Zhang, Guo, Yu, Moshkovich, Wehrstedt, Khabsa, Avalani, Bhatt, Mankus, Hasson, Lennie, Reso, Groshev, Naumov, Lathi, Keneally, Liu, Seltzer, Valko, Restrepo, Patel, Vyatskov, Samvelyan, Clark, Macey, Wang, Hermoso, Metanat, Rastegari, Bansal, Santhanam, Parks, White, Bawa, Singhal, Egebo, Usunier, Mehta, Laptev, Dong, Cheng, Chernoguz, Hart, Salpekar, Kalinli, Kent, Parekh, Saab, Balaji, Rittner, Bontrager, Roux, Dollar, Zvyagina, Ratanchandani, Yuvraj, Liang, Alao, Rodriguez, Ayub, Murthy, Nayani, Mitra, Parthasarathy, Li, Hogan, Battey, Wang, Howes, Rinott, Mehta, Siby, Bondu, Datta, Chugh, Hunt, Dhillon, Sidorov, Pan, Mahajan, Verma, Yamamoto, Ramaswamy, Lindsay, Lindsay, Feng, Lin, Zha, Patil, Shankar, Zhang, Zhang, Wang, Agarwal, Sajuyigbe, Chintala, Max, Chen, Kehoe, Satterfield, Govindaprasad, Gupta, Deng, Cho, Virk, Subramanian, Choudhury,
  Goldman, Remez, Glaser, Best, Koehler, Robinson, Li, Zhang, Matthews, Chou, Shaked, Vontimitta, Ajayi, Montanez, Mohan, Kumar, Mangla, Ionescu, Poenaru, Mihailescu, Ivanov, Li, Wang, Jiang, Bouaziz, Constable, Tang, Wu, Wang, Wu, Gao, Kleinman, Chen, Hu, Jia, Qi, Li, Zhang, Zhang, Adi, Nam, Yu, Wang, Zhao, Hao, Qian, Li, He, Rait, DeVito, Rosnbrick, Wen, Yang, Zhao, and Ma}]{grattafiori2024llama3herdmodels}
Aaron Grattafiori, Abhimanyu Dubey, Abhinav Jauhri, Abhinav Pandey, Abhishek Kadian, Ahmad Al-Dahle, Aiesha Letman, Akhil Mathur, Alan Schelten, Alex Vaughan, Amy Yang, Angela Fan, Anirudh Goyal, Anthony Hartshorn, Aobo Yang, Archi Mitra, Archie Sravankumar, Artem Korenev, Arthur Hinsvark, and 542 others. 2024.
\newblock \href {https://arxiv.org/abs/2407.21783} {The llama 3 herd of models}.
\newblock \emph{Preprint}, arXiv:2407.21783.

\bibitem[{Greenblatt et~al.(2024{\natexlab{a}})Greenblatt, Denison, Wright, Roger, MacDiarmid, Marks, Treutlein, Belonax, Chen, Duvenaud et~al.}]{greenblatt2024alignment}
Ryan Greenblatt, Carson Denison, Benjamin Wright, Fabien Roger, Monte MacDiarmid, Sam Marks, Johannes Treutlein, Tim Belonax, Jack Chen, David Duvenaud, and 1 others. 2024{\natexlab{a}}.
\newblock Alignment faking in large language models.
\newblock \emph{arXiv preprint arXiv:2412.14093}.

\bibitem[{Greenblatt et~al.(2024{\natexlab{b}})Greenblatt, Roger, Krasheninnikov, and Krueger}]{greenblatt2024stress}
Ryan Greenblatt, Fabien Roger, Dmitrii Krasheninnikov, and David Krueger. 2024{\natexlab{b}}.
\newblock Stress-testing capability elicitation with password-locked models.
\newblock \emph{Advances in Neural Information Processing Systems}, 37:69144--69175.

\bibitem[{Guo et~al.(2024)Guo, Zhu, Yang, Xie, Dong, Zhang, Chen, Bi, Wu, Li, Luo, Xiong, and Liang}]{guo2024deepseekcoderlargelanguagemodel}
Daya Guo, Qihao Zhu, Dejian Yang, Zhenda Xie, Kai Dong, Wentao Zhang, Guanting Chen, Xiao Bi, Y.~Wu, Y.~K. Li, Fuli Luo, Yingfei Xiong, and Wenfeng Liang. 2024.
\newblock \href {https://arxiv.org/abs/2401.14196} {Deepseek-coder: When the large language model meets programming -- the rise of code intelligence}.
\newblock \emph{Preprint}, arXiv:2401.14196.

\bibitem[{Hendrycks et~al.(2021{\natexlab{a}})Hendrycks, Burns, Basart, Zou, Mazeika, Song, and Steinhardt}]{hendrycks2021measuring}
Dan Hendrycks, Collin Burns, Steven Basart, Andy Zou, Mantas Mazeika, Dawn Song, and Jacob Steinhardt. 2021{\natexlab{a}}.
\newblock \href {https://openreview.net/forum?id=d7KBjmI3GmQ} {Measuring massive multitask language understanding}.
\newblock In \emph{International Conference on Learning Representations}.

\bibitem[{Hendrycks et~al.(2021{\natexlab{b}})Hendrycks, Burns, Kadavath, Arora, Basart, Tang, Song, and Steinhardt}]{hendrycks2021measuringmath}
Dan Hendrycks, Collin Burns, Saurav Kadavath, Akul Arora, Steven Basart, Eric Tang, Dawn Song, and Jacob Steinhardt. 2021{\natexlab{b}}.
\newblock \href {https://openreview.net/forum?id=7Bywt2mQsCe} {Measuring mathematical problem solving with the {MATH} dataset}.
\newblock In \emph{Thirty-fifth Conference on Neural Information Processing Systems Datasets and Benchmarks Track (Round 2)}.

\bibitem[{Hofst{\"a}tter et~al.(2025)Hofst{\"a}tter, van~der Weij, Teoh, Djoneva, Bartsch, and Ward}]{hofstatter2025the}
Felix Hofst{\"a}tter, Teun van~der Weij, Jayden Teoh, Rada Djoneva, Henning Bartsch, and Francis~Rhys Ward. 2025.
\newblock \href {https://openreview.net/forum?id=kT0EVqL77E} {The elicitation game: Evaluating capability elicitation techniques}.
\newblock In \emph{Forty-second International Conference on Machine Learning}.

\bibitem[{Hu et~al.(2022)Hu, yelong shen, Wallis, Allen-Zhu, Li, Wang, Wang, and Chen}]{hu2022lora}
Edward~J Hu, yelong shen, Phillip Wallis, Zeyuan Allen-Zhu, Yuanzhi Li, Shean Wang, Lu~Wang, and Weizhu Chen. 2022.
\newblock \href {https://openreview.net/forum?id=nZeVKeeFYf9} {Lo{RA}: Low-rank adaptation of large language models}.
\newblock In \emph{International Conference on Learning Representations}.

\bibitem[{Huang et~al.(2023)Huang, Zhao, Backes, Shen, and Zhang}]{huang2023composite}
Hai Huang, Zhengyu Zhao, Michael Backes, Yun Shen, and Yang Zhang. 2023.
\newblock Composite backdoor attacks against large language models.
\newblock \emph{arXiv preprint arXiv:2310.07676}.

\bibitem[{Hubinger et~al.(2024)Hubinger, Denison, Mu, Lambert, Tong, MacDiarmid, Lanham, Ziegler, Maxwell, Cheng et~al.}]{hubinger2024sleeper}
Evan Hubinger, Carson Denison, Jesse Mu, Mike Lambert, Meg Tong, Monte MacDiarmid, Tamera Lanham, Daniel~M Ziegler, Tim Maxwell, Newton Cheng, and 1 others. 2024.
\newblock Sleeper agents: Training deceptive llms that persist through safety training.
\newblock \emph{arXiv preprint arXiv:2401.05566}.

\bibitem[{Ilharco et~al.(2023)Ilharco, Ribeiro, Wortsman, Schmidt, Hajishirzi, and Farhadi}]{ilharco2023editing}
Gabriel Ilharco, Marco~Tulio Ribeiro, Mitchell Wortsman, Ludwig Schmidt, Hannaneh Hajishirzi, and Ali Farhadi. 2023.
\newblock \href {https://openreview.net/forum?id=6t0Kwf8-jrj} {Editing models with task arithmetic}.
\newblock In \emph{The Eleventh International Conference on Learning Representations}.

\bibitem[{Kalajdzievski(2023)}]{kalajdzievski2023rankstabilizationscalingfactor}
Damjan Kalajdzievski. 2023.
\newblock \href {https://arxiv.org/abs/2312.03732} {A rank stabilization scaling factor for fine-tuning with lora}.
\newblock \emph{Preprint}, arXiv:2312.03732.

\bibitem[{Kaur and Verma(2023)}]{KaurAdaptive}
Amardeep Kaur and Amandeep Verma. 2023.
\newblock \href {https://doi.org/10.1155/2023/3922393} {Adaptive access control mechanism (aacm) for enterprise cloud computing}.
\newblock \emph{Journal of Electrical and Computer Engineering}, 2023(1):3922393.

\bibitem[{Kotha et~al.(2024)Kotha, Springer, and Raghunathan}]{kotha2024understanding}
Suhas Kotha, Jacob~Mitchell Springer, and Aditi Raghunathan. 2024.
\newblock \href {https://openreview.net/forum?id=VrHiF2hsrm} {Understanding catastrophic forgetting in language models via implicit inference}.
\newblock In \emph{The Twelfth International Conference on Learning Representations}.

\bibitem[{Kwon et~al.(2023)Kwon, Li, Zhuang, Sheng, Zheng, Yu, Gonzalez, Zhang, and Stoica}]{kwon2023efficient}
Woosuk Kwon, Zhuohan Li, Siyuan Zhuang, Ying Sheng, Lianmin Zheng, Cody~Hao Yu, Joseph~E. Gonzalez, Hao Zhang, and Ion Stoica. 2023.
\newblock Efficient memory management for large language model serving with pagedattention.
\newblock In \emph{Proceedings of the ACM SIGOPS 29th Symposium on Operating Systems Principles}.

\bibitem[{Lee et~al.(2025)Lee, Ko, Pedapati, Chung, Yeh, and Chen}]{lee2025star}
Yu-Ang Lee, Ching-Yun Ko, Tejaswini Pedapati, I-Hsin Chung, Mi-Yen Yeh, and Pin-Yu Chen. 2025.
\newblock Star: Spectral truncation and rescale for model merging.
\newblock In \emph{Proceedings of the 2025 Conference of the Nations of the Americas Chapter of the Association for Computational Linguistics: Human Language Technologies (Volume 2: Short Papers)}, pages 496--505.

\bibitem[{Li et~al.(2024)Li, Huang, Zhao, Ma, and Sun}]{li2024backdoorllm}
Yige Li, Hanxun Huang, Yunhan Zhao, Xingjun Ma, and Jun Sun. 2024.
\newblock Backdoorllm: A comprehensive benchmark for backdoor attacks and defenses on large language models.
\newblock \emph{arXiv preprint arXiv:2408.12798}.

\bibitem[{Lin(2004)}]{lin2004rouge}
Chin-Yew Lin. 2004.
\newblock Rouge: A package for automatic evaluation of summaries.
\newblock In \emph{Text summarization branches out}, pages 74--81.

\bibitem[{Liu et~al.(2025)Liu, Li, Suh, Vorobeychik, Mao, Jha, McDaniel, Sun, Li, and Xiao}]{liu2025autodanturbo}
Xiaogeng Liu, Peiran Li, G.~Edward Suh, Yevgeniy Vorobeychik, Zhuoqing Mao, Somesh Jha, Patrick McDaniel, Huan Sun, Bo~Li, and Chaowei Xiao. 2025.
\newblock \href {https://openreview.net/forum?id=bhK7U37VW8} {Auto{DAN}-turbo: A lifelong agent for strategy self-exploration to jailbreak {LLM}s}.
\newblock In \emph{The Thirteenth International Conference on Learning Representations}.

\bibitem[{Lundy et~al.(2024)Lundy, Raman, Fu, and Leyton-Brown}]{lundy2024pay}
Taylor Lundy, Narun Raman, Hu~Fu, and Kevin Leyton-Brown. 2024.
\newblock Pay to (not) play: monetizing impatience in mobile games.
\newblock In \emph{Proceedings of the AAAI Conference on Artificial Intelligence}, volume~38, pages 9856--9864.

\bibitem[{Mangrulkar et~al.(2022)Mangrulkar, Gugger, Debut, Belkada, Paul, Bossan, and Tietz}]{peft}
Sourab Mangrulkar, Sylvain Gugger, Lysandre Debut, Younes Belkada, Sayak Paul, Benjamin Bossan, and Marian Tietz. 2022.
\newblock {PEFT}: State-of-the-art parameter-efficient fine-tuning methods.
\newblock \url{https://github.com/huggingface/peft}.

\bibitem[{Mehrotra et~al.(2024)Mehrotra, Zampetakis, Kassianik, Nelson, Anderson, Singer, and Karbasi}]{mehrotra2024tree}
Anay Mehrotra, Manolis Zampetakis, Paul Kassianik, Blaine Nelson, Hyrum~S Anderson, Yaron Singer, and Amin Karbasi. 2024.
\newblock \href {https://openreview.net/forum?id=SoM3vngOH5} {Tree of attacks: Jailbreaking black-box {LLM}s automatically}.
\newblock In \emph{The Thirty-eighth Annual Conference on Neural Information Processing Systems}.

\bibitem[{Ong et~al.(2024)Ong, Almahairi, Wu, Chiang, Wu, Gonzalez, Kadous, and Stoica}]{ong2024routellm}
Isaac Ong, Amjad Almahairi, Vincent Wu, Wei-Lin Chiang, Tianhao Wu, Joseph~E Gonzalez, M~Waleed Kadous, and Ion Stoica. 2024.
\newblock Routellm: Learning to route llms with preference data.
\newblock \emph{arXiv preprint arXiv:2406.18665}.

\bibitem[{Ortiz-Jimenez et~al.(2023)Ortiz-Jimenez, Favero, and Frossard}]{ortiz-jimenez2023task}
Guillermo Ortiz-Jimenez, Alessandro Favero, and Pascal Frossard. 2023.
\newblock \href {https://openreview.net/forum?id=0A9f2jZDGW} {Task arithmetic in the tangent space: Improved editing of pre-trained models}.
\newblock In \emph{Thirty-seventh Conference on Neural Information Processing Systems}.

\bibitem[{Ouyang et~al.(2022)Ouyang, Wu, Jiang, Almeida, Wainwright, Mishkin, Zhang, Agarwal, Slama, Gray, Schulman, Hilton, Kelton, Miller, Simens, Askell, Welinder, Christiano, Leike, and Lowe}]{ouyang2022training}
Long Ouyang, Jeffrey Wu, Xu~Jiang, Diogo Almeida, Carroll Wainwright, Pamela Mishkin, Chong Zhang, Sandhini Agarwal, Katarina Slama, Alex Gray, John Schulman, Jacob Hilton, Fraser Kelton, Luke Miller, Maddie Simens, Amanda Askell, Peter Welinder, Paul Christiano, Jan Leike, and Ryan Lowe. 2022.
\newblock \href {https://openreview.net/forum?id=TG8KACxEON} {Training language models to follow instructions with human feedback}.
\newblock In \emph{Advances in Neural Information Processing Systems}.

\bibitem[{Prabhakar et~al.(2025)Prabhakar, Li, Narasimhan, Kakade, Malach, and Jelassi}]{prabhakar-etal-2025-lora}
Akshara Prabhakar, Yuanzhi Li, Karthik Narasimhan, Sham Kakade, Eran Malach, and Samy Jelassi. 2025.
\newblock \href {https://aclanthology.org/2025.coling-industry.55/} {{L}o{RA} soups: Merging {L}o{RA}s for practical skill composition tasks}.
\newblock In \emph{Proceedings of the 31st International Conference on Computational Linguistics: Industry Track}, pages 644--655, Abu Dhabi, UAE. Association for Computational Linguistics.

\bibitem[{Shao et~al.(2024)Shao, Wang, Zhu, Xu, Song, Bi, Zhang, Zhang, Li, Wu, and Guo}]{shao2024deepseekmathpushinglimitsmathematical}
Zhihong Shao, Peiyi Wang, Qihao Zhu, Runxin Xu, Junxiao Song, Xiao Bi, Haowei Zhang, Mingchuan Zhang, Y.~K. Li, Y.~Wu, and Daya Guo. 2024.
\newblock \href {https://arxiv.org/abs/2402.03300} {Deepseekmath: Pushing the limits of mathematical reasoning in open language models}.
\newblock \emph{Preprint}, arXiv:2402.03300.

\bibitem[{Shayegani et~al.(2024)Shayegani, Dong, and Abu-Ghazaleh}]{shayegani2024jailbreak}
Erfan Shayegani, Yue Dong, and Nael Abu-Ghazaleh. 2024.
\newblock \href {https://openreview.net/forum?id=plmBsXHxgR} {Jailbreak in pieces: Compositional adversarial attacks on multi-modal language models}.
\newblock In \emph{The Twelfth International Conference on Learning Representations}.

\bibitem[{Shayegani et~al.(2023)Shayegani, Mamun, Fu, Zaree, Dong, and Abu-Ghazaleh}]{shayegani2023surveyvulnerabilitieslargelanguage}
Erfan Shayegani, Md~Abdullah~Al Mamun, Yu~Fu, Pedram Zaree, Yue Dong, and Nael Abu-Ghazaleh. 2023.
\newblock \href {https://arxiv.org/abs/2310.10844} {Survey of vulnerabilities in large language models revealed by adversarial attacks}.
\newblock \emph{Preprint}, arXiv:2310.10844.

\bibitem[{Sheshadri et~al.(2025)Sheshadri, Ewart, Guo, Lynch, Wu, Hebbar, Sleight, Stickland, Perez, Hadfield-Menell, and Casper}]{sheshadri2025latent}
Abhay Sheshadri, Aidan Ewart, Phillip~Huang Guo, Aengus Lynch, Cindy Wu, Vivek Hebbar, Henry Sleight, Asa~Cooper Stickland, Ethan Perez, Dylan Hadfield-Menell, and Stephen Casper. 2025.
\newblock \href {https://openreview.net/forum?id=6LxMeRlkWl} {Latent adversarial training improves robustness to persistent harmful behaviors in {LLM}s}.
\newblock \emph{Transactions on Machine Learning Research}.

\bibitem[{Su et~al.(2025)Su, Gao, Ding, and Ma}]{su2025identity}
Hongyu Su, Yifeng Gao, Yifan Ding, and Xingjun Ma. 2025.
\newblock Identity lock: Locking api fine-tuned llms with identity-based wake words.
\newblock \emph{arXiv preprint arXiv:2503.10668}.

\bibitem[{Sutton et~al.(2025)Sutton, Zhou, Leete, Gorban, and Tyukin}]{sutton2025staining}
Oliver~J Sutton, Qinghua Zhou, George Leete, Alexander~N Gorban, and Ivan~Y Tyukin. 2025.
\newblock Staining and locking computer vision models without retraining.
\newblock \emph{arXiv preprint arXiv:2507.22000}.

\bibitem[{Tang et~al.(2024)Tang, Chuang, Cai, Du, and Hu}]{tangsecure}
Ruixiang Tang, Yu-Neng Chuang, Xuanting Cai, Mengnan Du, and Xia Hu. 2024.
\newblock \href {https://doi.org/10.18653/v1/2024.findings-naacl.256} {Secure your model: An effective key prompt protection mechanism for large language models}.
\newblock In \emph{Findings of the Association for Computational Linguistics: NAACL 2024}, pages 4061--4073, Mexico City, Mexico. Association for Computational Linguistics.

\bibitem[{Yadav et~al.(2023)Yadav, Tam, Choshen, Raffel, and Bansal}]{yadav2023tiesmerging}
Prateek Yadav, Derek Tam, Leshem Choshen, Colin Raffel, and Mohit Bansal. 2023.
\newblock \href {https://openreview.net/forum?id=xtaX3WyCj1} {{TIES}-merging: Resolving interference when merging models}.
\newblock In \emph{Thirty-seventh Conference on Neural Information Processing Systems}.

\bibitem[{Yan et~al.(2025)Yan, Li, Wang, Chen, He, and Li}]{embedX}
Nan Yan, Yuqing Li, Xiong Wang, Jing Chen, Kun He, and Bo~Li. 2025.
\newblock Embedx: embedding-based cross-trigger backdoor attack against large language models.
\newblock In \emph{Proceedings of the 34th USENIX Conference on Security Symposium}, SEC '25, USA. USENIX Association.

\bibitem[{Yu et~al.(2018)Yu, Zhang, Yang, Yasunaga, Wang, Li, Ma, Li, Yao, Roman et~al.}]{yu2018spider}
Tao Yu, Rui Zhang, Kai Yang, Michihiro Yasunaga, Dongxu Wang, Zifan Li, James Ma, Irene Li, Qingning Yao, Shanelle Roman, and 1 others. 2018.
\newblock Spider: A large-scale human-labeled dataset for complex and cross-domain semantic parsing and text-to-sql task.
\newblock \emph{arXiv preprint arXiv:1809.08887}.

\bibitem[{Zeng and Lu(2022)}]{zengunsupervised}
Guangtao Zeng and Wei Lu. 2022.
\newblock \href {https://doi.org/10.18653/v1/2022.emnlp-main.685} {Unsupervised non-transferable text classification}.
\newblock In \emph{Proceedings of the 2022 Conference on Empirical Methods in Natural Language Processing}, pages 10071--10084, Abu Dhabi, United Arab Emirates. Association for Computational Linguistics.

\bibitem[{Zhang et~al.(2025{\natexlab{a}})Zhang, Chen, He, Lou, Li, Feng, Song, Liu, Ren, and Yang}]{quada}
Jiawen Zhang, Kejia Chen, Lipeng He, Jian Lou, Dan Li, Zunlei Feng, Mingli Song, Jian Liu, Kui Ren, and Xiaohu Yang. 2025{\natexlab{a}}.
\newblock Activation approximations can incur safety vulnerabilities even in aligned llms: comprehensive analysis and defense.
\newblock In \emph{Proceedings of the 34th USENIX Conference on Security Symposium}, SEC '25, USA. USENIX Association.

\bibitem[{Zhang et~al.(2024{\natexlab{a}})Zhang, Lin, Bai, and Mei}]{zhang2024negative}
Ruiqi Zhang, Licong Lin, Yu~Bai, and Song Mei. 2024{\natexlab{a}}.
\newblock \href {https://openreview.net/forum?id=MXLBXjQkmb} {Negative preference optimization: From catastrophic collapse to effective unlearning}.
\newblock In \emph{First Conference on Language Modeling}.

\bibitem[{Zhang et~al.(2024{\natexlab{b}})Zhang, Carlini, and Ippolito}]{zhang2024effective}
Yiming Zhang, Nicholas Carlini, and Daphne Ippolito. 2024{\natexlab{b}}.
\newblock \href {https://openreview.net/forum?id=0o95CVdNuz} {Effective prompt extraction from language models}.
\newblock In \emph{First Conference on Language Modeling}.

\bibitem[{Zhang et~al.(2025{\natexlab{b}})Zhang, Rando, Evtimov, Chi, Smith, Carlini, Tram{\`e}r, and Ippolito}]{zhang2025persistent}
Yiming Zhang, Javier Rando, Ivan Evtimov, Jianfeng Chi, Eric~Michael Smith, Nicholas Carlini, Florian Tram{\`e}r, and Daphne Ippolito. 2025{\natexlab{b}}.
\newblock \href {https://openreview.net/forum?id=eiqrnVaeIw} {Persistent pre-training poisoning of {LLM}s}.
\newblock In \emph{The Thirteenth International Conference on Learning Representations}.

\bibitem[{Zhong et~al.(2017)Zhong, Xiong, and Socher}]{zhongSeq2SQL2017}
Victor Zhong, Caiming Xiong, and Richard Socher. 2017.
\newblock Seq2sql: Generating structured queries from natural language using reinforcement learning.
\newblock \emph{CoRR}, abs/1709.00103.

\bibitem[{Ziegler et~al.(2022)Ziegler, Nix, Chan, Bauman, Schmidt-Nielsen, Lin, Scherlis, Nabeshima, Weinstein-Raun, de~Haas, Shlegeris, and Thomas}]{ziegler2022adversarial}
Daniel Ziegler, Seraphina Nix, Lawrence Chan, Tim Bauman, Peter Schmidt-Nielsen, Tao Lin, Adam Scherlis, Noa Nabeshima, Benjamin Weinstein-Raun, Daniel de~Haas, Buck Shlegeris, and Nate Thomas. 2022.
\newblock \href {https://openreview.net/forum?id=NtJyGXo0nF} {Adversarial training for high-stakes reliability}.
\newblock In \emph{Advances in Neural Information Processing Systems}.

\bibitem[{Zou et~al.(2024)Zou, Phan, Wang, Duenas, Lin, Andriushchenko, Kolter, Fredrikson, and Hendrycks}]{zou2024improving}
Andy Zou, Long Phan, Justin Wang, Derek Duenas, Maxwell Lin, Maksym Andriushchenko, J~Zico Kolter, Matt Fredrikson, and Dan Hendrycks. 2024.
\newblock \href {https://openreview.net/forum?id=IbIB8SBKFV} {Improving alignment and robustness with circuit breakers}.
\newblock In \emph{The Thirty-eighth Annual Conference on Neural Information Processing Systems}.

\bibitem[{Zou et~al.(2023)Zou, Wang, Carlini, Nasr, Kolter, and Fredrikson}]{zou2023gcg}
Andy Zou, Zifan Wang, Nicholas Carlini, Milad Nasr, J.~Zico Kolter, and Matt Fredrikson. 2023.
\newblock \href {https://arxiv.org/abs/2307.15043} {Universal and transferable adversarial attacks on aligned language models}.
\newblock \emph{Preprint}, arXiv:2307.15043.

\end{thebibliography}
